\documentclass[traditabstract,longauth]{aa}
\usepackage{graphicx}
\usepackage{lscape}
\usepackage{amsmath}
\usepackage{amssymb}

\usepackage{txfonts}
\usepackage{longtable,lscape}
\usepackage{natbib}
\usepackage{longtable}
\bibpunct{(}{)}{;}{a}{}{,} 

\newcommand{\corot}{\it{CoRoT}\rm}  
\def\vsini{$v$\,sin\,$i$}             
\def\ms{\hbox{\,m\,s$^{-1}$}}         
\def\m2s2{\hbox{\,m$^{2}$\,s$^{-2}$}} 
\def\kms{\hbox{\,km\,s$^{-1}$}}       
\def\gcm3{\hbox{\,g\,cm$^{-3}$}}      
\def\vsini{\hbox{$v$\,sin\,$i_\star$}}      
\def\Msun{\hbox{$M_{\odot}$}}         
\def\Rsun{\hbox{$R_{\odot}$}}
\def\Mjup{\hbox{$\mathrm{M}_{\rm Jup}$}}
\def\Rjup{\hbox{$\mathrm{R}_{\rm Jup}$}}
\def\Mearth{\hbox{$\mathrm{M}_{\oplus}$}}
\def\Rearth{\hbox{$\mathrm{R}_{\oplus}$}}

\def\chisq{\mbox{$\chi^2$}}

\def\target{CoRoT-24}
\def\targetb{CoRoT-24b}
\def\targetc{CoRoT-24c}
\def\kepler{\it{Kepler}\rm}
\usepackage{color}

\begin{document}
\title{Transiting exoplanets from the CoRoT space mission\thanks{The CoRoT space mission, launched on December 27th 2006, has been developed and is operated by CNES, with the contribution of Austria, Belgium, Brazil , ESA (RSSD and Science Program), Germany, and Spain. Some of the observations were made with the HARPS spectrograph at ESO La Silla Observatory (184.C-0639) and with the HIRES spectrograph at the Keck telescope (N035Hr, N143Hr 260 and N095Hr). Partly based on observations obtained at ESO Paranal Observatory, Chile (086.C-0235(A) and B).}}
\subtitle{XXVI. \target : A transiting multiplanet system. }

\author{R.~Alonso\inst{1,2,3}
\and C.~Moutou\inst{4}
\and M.~Endl\inst{5}
\and J.-M.~Almenara\inst{4}
\and E.W.~Guenther\inst{6,19}
\and M.~Deleuil\inst{4}
\and A.~Hatzes\inst{6}
\and S.~Aigrain\inst{7}
\and M.~Auvergne\inst{8}
\and A.~Baglin\inst{8}
\and P.~Barge\inst{4}
\and A.~S.~Bonomo\inst{4}
\and P.~Bord\'e\inst{9}
\and F.~Bouchy\inst{10,11,1}
\and C.~Cavarroc\inst{9}
\and J.~Cabrera\inst{12}
\and S.~Carpano\inst{13}
\and Sz.~Csizmadia\inst{12}
\and W.~D.~Cochran\inst{5}
\and H.~J.~Deeg\inst{2,3}
\and R.~F.~D\'\i az\inst{4,1}
\and R.~Dvorak\inst{14}
\and A.~Erikson\inst{12}
\and S.~Ferraz-Mello\inst{15}
\and M.~Fridlund\inst{12,13}
\and T.~Fruth\inst{12,25}
\and D.~Gandolfi\inst{26}
\and M.~Gillon\inst{16}
\and S.~Grziwa\inst{17}
\and T.~Guillot\inst{18}
\and G.~H\'ebrard\inst{10,11}
\and L.~Jorda\inst{4}
\and A.~L\'eger\inst{9}
\and H.~Lammer\inst{19}
\and C.~Lovis\inst{1}
\and P.~J.~MacQueen\inst{5}
\and T.~Mazeh\inst{20}
\and A.~Ofir\inst{21}
\and M.~Ollivier\inst{9}
\and T.~Pasternacki\inst{12}
\and M.~P\"atzold\inst{17}
\and D.~Queloz\inst{1}
\and H.~Rauer\inst{12,22}
\and D.~Rouan\inst{8}
\and A.~Santerne\inst{4,23}
\and J.~Schneider\inst{24}
\and M.~Tadeu dos Santos\inst{15}
\and B.~Tingley\inst{2,3,27}
\and R.~Titz-Weider\inst{12}
\and J.~Weingrill\inst{19}
\and G.~Wuchterl\inst{6}
}

\institute{
Observatoire de l'Universit\'e de Gen\`eve, 51 chemin des Maillettes, 1290 Sauverny, Switzerland\label{Geneve}
\and Instituto de Astrof{\'i}sica de Canarias, E-38205 La Laguna, Tenerife, Spain\label{IAC}
\and Dpto. de Astrof\'isica, Universidad de La Laguna, 38206 La Laguna, Tenerife, Spain\label{La Laguna}
\and Laboratoire d'Astrophysique de Marseille, CNRS \& Aix-Marseille University, 38 rue Fr\'ed\'eric Joliot-Curie, 13388 Marseille cedex 13, France\label{LAM}
\and McDonald Observatory, University of Texas at Austin, Austin, TX 78712, USA\label{McD}
\and Th\"uringer Landessternwarte, Sternwarte 5, Tautenburg, D-07778 Tautenburg, Germany\label{Tautenburg}
\and Oxford Astrophyiscs, Denys Wilkinson Building, Keble Road, Oxford OX1 3RH\label{Oxford}
\and LESIA, Observatoire de Paris, Place Jules Janssen, 92195 Meudon cedex, France\label{LESIA}
\and Institut d'Astrophysique Spatiale, Universit\'e Paris-Sud \& CNRS, 91405 Orsay, France\label{IAS}
\and Observatoire de Haute Provence, 04670 Saint Michel l'Observatoire, France\label{OHP}
\and Institut d'Astrophysique de Paris, UMR7095 CNRS, Universit\'e Pierre \& Marie Curie, 98bis boulevard Arago, 75014 Paris, France\label{IAP}
\and Institute of Planetary Research, German Aerospace Center, Rutherfordstrasse 2, 12489 Berlin, Germany\label{DLR}
\and Research and Scientific Support Department, ESTEC/ESA, PO Box 299, 2200 AG Noordwijk, The Netherlands\label{ESA}
\and University of Vienna, Institute of Astronomy, T\"urkenschanzstr. 17, A-1180 Vienna, Austria\label{Wien}
\and IAG, University of S\~ao Paulo, Brasil\label{Brasil}
\and University of Li\`ege, All\'ee du 6 ao\^ut 17, Sart Tilman, Li\`ege 1, Belgium\label{Liege}
\and Rheinisches Institut f\"ur Umweltforschung, Abteilung Planetenforschung, an der Universit\"at zu K\"oln, Aachener Strasse 209, 50931, Germany
\and Universit\'e de Nice-Sophia Antipolis, CNRS UMR 6202, Observatoire de la C\^ote d'Azur, BP 4229, 06304 Nice Cedex 4, France\label{OCA}
\and Space Research Institute, Austrian Academy of Science, Schmiedlstr. 6, A-8042 Graz, Austria\label{Graz}
\and School of Physics and Astronomy, Raymond and Beverly Sackler Faculty of Exact Sciences, Tel Aviv University, Tel Aviv, Israel\label{Tel Aviv}
\and Institut f\"ur Astrophysik, Georg-August-Universit\"at, Friedrich-Hund-Platz 1, 37077 G\"ottingen, Germany \label{ifa}
\and Center for Astronomy and Astrophysics, TU Berlin, Hardenbergstr. 36, 10623 Berlin, Germany\label{ZAA}
\and Centro de Astrof\'isica, Universidade do Porto, Rua das Estrelas, 4150-762 Porto, Portugal
\and LUTH, Observatoire de Paris, CNRS, Universit\'e Paris Diderot; 5 place Jules Janssen, 92195 Meudon, France\label{LUTh}
\and{German Space Operations Center, German Aerospace Center, M\"unchner Strasse 20, 82234 We\ss ling, Germany\label{GAC}
\and{Landessternwarte K\"onigstuhl, Zentrum f\"ur Astronomie der Universit\"at Heidelberg, K\"onigstuhl 12, D-69117 Heidelberg, Germany}
\and{Institut for Fysik og Astronomi, Aarhus Universitet, Ny Munkegade 120, 8000 Aarhus C, Denmark}
}
}

   \date{Received ...; accepted ...}
  \abstract
    {We present the discovery of a candidate multiply transiting system, the first one found in the \corot\ mission. Two transit-like features with periods of 5.11 and 11.76~d are detected in the \corot\ light curve around a main sequence K1V star of $r$=15.1. If the features are due to transiting planets around the same star, these would correspond to objects of 3.7$\pm$0.4 and 5.0$\pm$0.5 \Rearth\ , respectively. Several radial velocities serve to provide an upper limit of 5.7~\Mearth\ for the 5.11~d signal and to tentatively measure a mass of 28$^{+11}_{-11}$~\Mearth\ for the object transiting with a 11.76~d period. These measurements imply low density objects, with a significant gaseous envelope. The detailed analysis of the photometric and spectroscopic data serve to estimate the probability that the observations are caused by transiting Neptune-sized planets as $>$26$\times$ higher than a blend scenario involving only one transiting planet and $>$900$\times$ higher than a scenario involving two blends and no planets. The radial velocities show a long-term modulation that might be attributed to a 1.5~\Mjup\ planet orbiting at 1.8~A.U. from the host, but more data are required to determine the precise orbital parameters of this companion.}

   \maketitle
\newcommand{\ud}{\,\mathrm{d}}
%

\section{Introduction}
\label{sec: intro}

With the exquisite photometric precision of the space missions, the field of exoplanets has reached several milestones in recent years. To enumerate a few, transiting planets with moderate temperatures like CoRoT-9b \citep{deeg2010}, with periods of several months, are key objects for studying the effect of the incoming flux from the host stars on their atmospheres. The detection of transit time variations was achieved in the system Kepler-9 \citep{holman2010}, and they have been used to measure the masses of several of the transiting planets found with the \kepler\ mission (e.g., \citealt{lissauer2011}, \citealt{fabrycky12}, \citealt{steffen13}). All the currently known transiting planets with densities compatible with rocky compositions were detected from space (e.g., CoRoT-7b, \citealt{leger2009}, Kepler-10b, \citealt{batalha2011}, 55-Cnc e, \citealt{winn2011,demory2011}), because their shallow transits -of a few hundred ppm- are a challenge for ground-based observations.
  
While exoplanetary systems have been routinely discovered or confirmed with the radial velocity method, the improvements in photometric precision and transit detection capabilities with the space missions \corot\ \citep{baglin06} and \kepler\ \citep{borucki10} allow for detection of transit-like features equivalent to objects with the size of an Earth and smaller (e.g., \citealt{barclay13}). The weakest transit signals demand considerable resources, as shown, for instance, with the intense effort needed to measure the mass of CoRoT-7b \citep{queloz2009}. Not only is the radial-velocity amplitude smaller, but the relative importance of the contribution by stellar activity is also higher, and new techniques need to be used to correct for this. This claim is supported by the more than six different analyses that have been published using the same radial velocity data set. When the transits are found around fainter host stars and/or with longer orbital periods, the current spectrographs do not reach the precision needed to measure the masses of the planets. In these cases, a study of the probabilities of the different scenarios that are compatible with all the follow-up observations performed on the target can result in a \emph{validation} of a planet candidate. This is the approach started by \cite{torres2004} and used in recent \kepler\ discovery papers (e.g., \citealt{fressin2011},\citealt{cochran2011}, \citealt{ballard2011}, \citealt{lissauer14}, \citealt{rowe14}). For small planets, it is also more difficult to exclude background eclipsing binaries, since fainter background stars could mimic the transit signal. 

This paper presents the discovery of the first multiply transiting system found by \corot  and, as such, is the first one that can be observed from the southern hemisphere observatories. We describe the \corot\ data in the next section. After the analysis, in Section~\ref{sec: analy_lc}, of the two transit signals that appear in the data, we describe the follow-up observations in Sections ~\ref{sec: groun_phot},~\ref{sec: rv}, and~\ref{sec: spec_ana}. A study of the multiple stellar systems that are compatible with the data is presented in Section~\ref{sec: simul}, and the probabilities of these systems are discussed in Section~\ref{sec: disc}.

\section{CoRoT observations}
\label{sec: observations}

The target star \target\ was observed on the detector E1 of the \corot\ satellite during its second \emph{Long Run} pointing at the direction opposite to the approximate galactic center (``anticenter" in the \corot\ jargon). The observations lasted from Nov 16, 2008 until March 8, 2009, covering a total span of 111.64 days. The target is faint, having a $r$ mag of 15.1 (see Table~\ref{startable}). A transit signal with a period of 11.76~d was detected in the framework of the \emph{alarm mode} \citep{Surace2008} close to the end of the observations, and thus the last 10.7 days of the run are sampled with a 32~s cadence, instead of the nominal 512~s of the rest of the observations. The $EN2$ data product\footnote{publicly available at {http://idoc-corot.ias.u-psud.fr}} consists of 45593 data points, out of which 28672 are sampled with a 32~s cadence. The aperture used on-board is a monochromatic one, since the telemetry only allows for a recovery of  chromatic information on a limited number of targets. Owing to the passage of the satellite over the south atlantic anomaly (SAA) and other instrumental effects, about 15\% of the data are flagged as bad data points. A detailed description of the performance and pipeline of \corot\ can be found in \cite{Auvergne2008}.

\begin{figure}
\centering
\includegraphics[width=9cm]{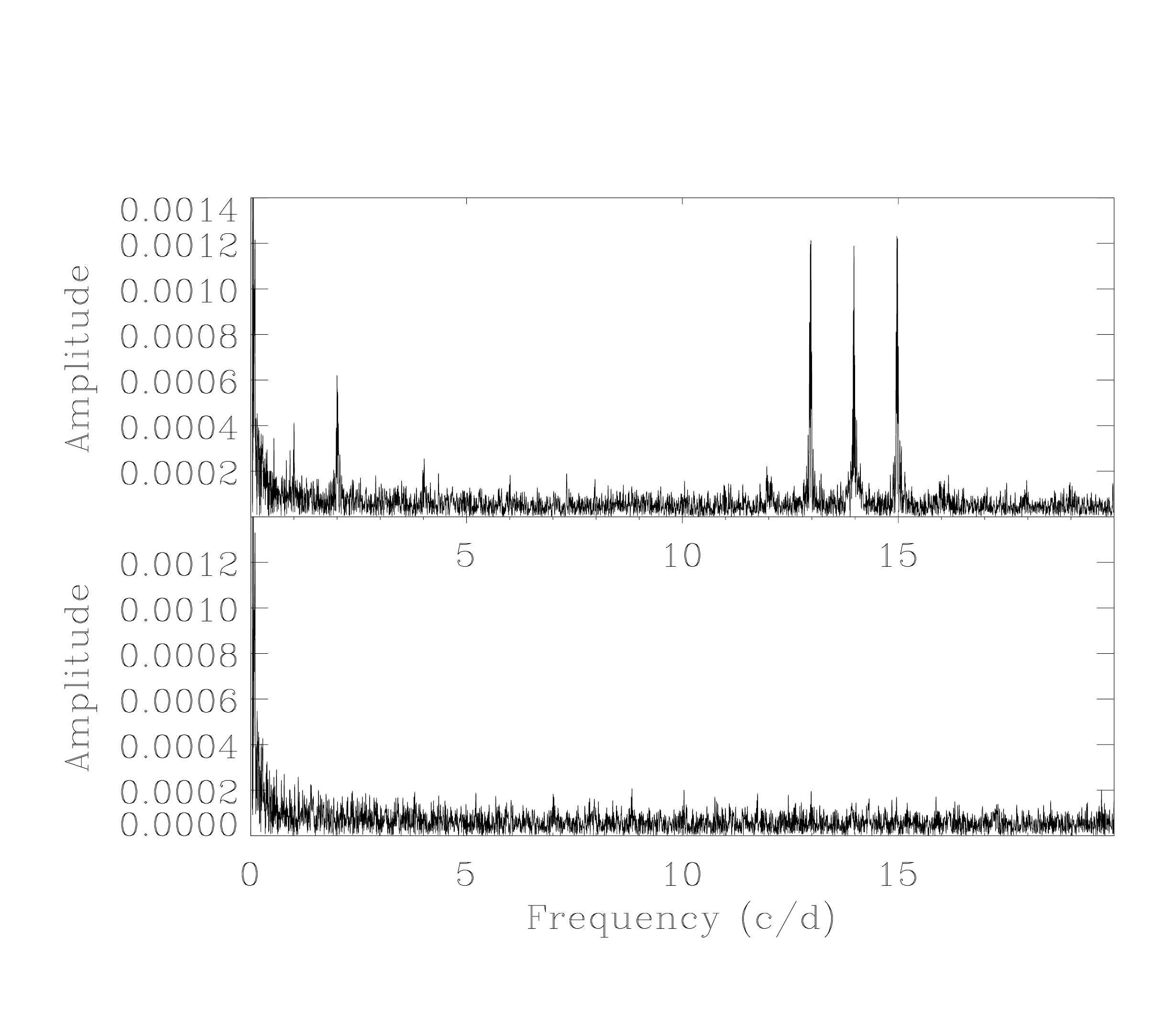}
\caption{Amplitude spectrum of the light curve, showing peaks at the frequencies of the satellite's orbital signal and the daily harmonics, before and after the correction described in Sect.~\ref{sec: observations}. }
\label{fig_spectrum}
\end{figure}

\begin{figure*}
\centering
\includegraphics[width=\textwidth]{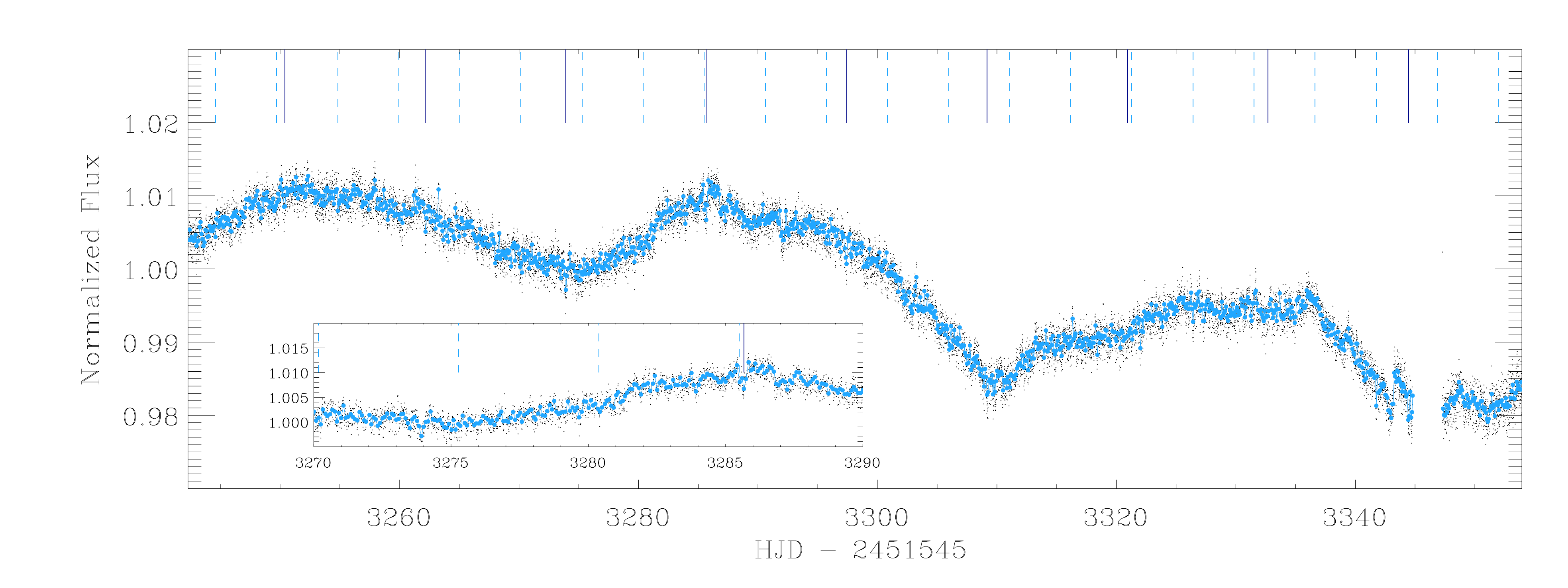}
\caption{Filtered light curve of \target\ . The black dots are the 512~s sampled data, while the blue curve has been combined in 1~h bins only for display purposes. The inset shows a zoom, where individual transits can be barely detected. The vertical lines show the positions of the transits of \targetb\ (dashed light blue) and \targetc\ (solid dark blue). }
\label{fig_lc}
\end{figure*}

As in previous analyses, the power spectrum of the curve without the data points flagged as invalid shows several peaks at the satellite orbital period and its daily aliases. The period of the satellite orbit is about 1.7~h, and thus it might affect the determinations of the transit parameters, because the time scales are comparable. To filter out this signal, we corrected each orbit $i$ with a model that was estimated by computing a Savitzky-Golay smoothed version of the 30 closest orbits to $i$, normalized independently \citep{Alonso2008}. While this process performs satisfactorily in removing the peak in the power spectrum at the orbital frequency, the daily aliases remain as a consequence of the gaps introduced in the data by removing the points that were flagged as bad. To avoid this side effect, we interpolated the data in the gaps by the use of an inpainting technique \citep{Sato2010}. The power spectrum before and after this correction is plotted in Fig.~\ref{fig_spectrum} and the filtered light curve is plotted in Fig.~\ref{fig_lc}.     

\section{Analysis of the CoRoT light curve}
\label{sec: analy_lc}

\begin{figure}[h]
\centering
\includegraphics[width=8.5cm]{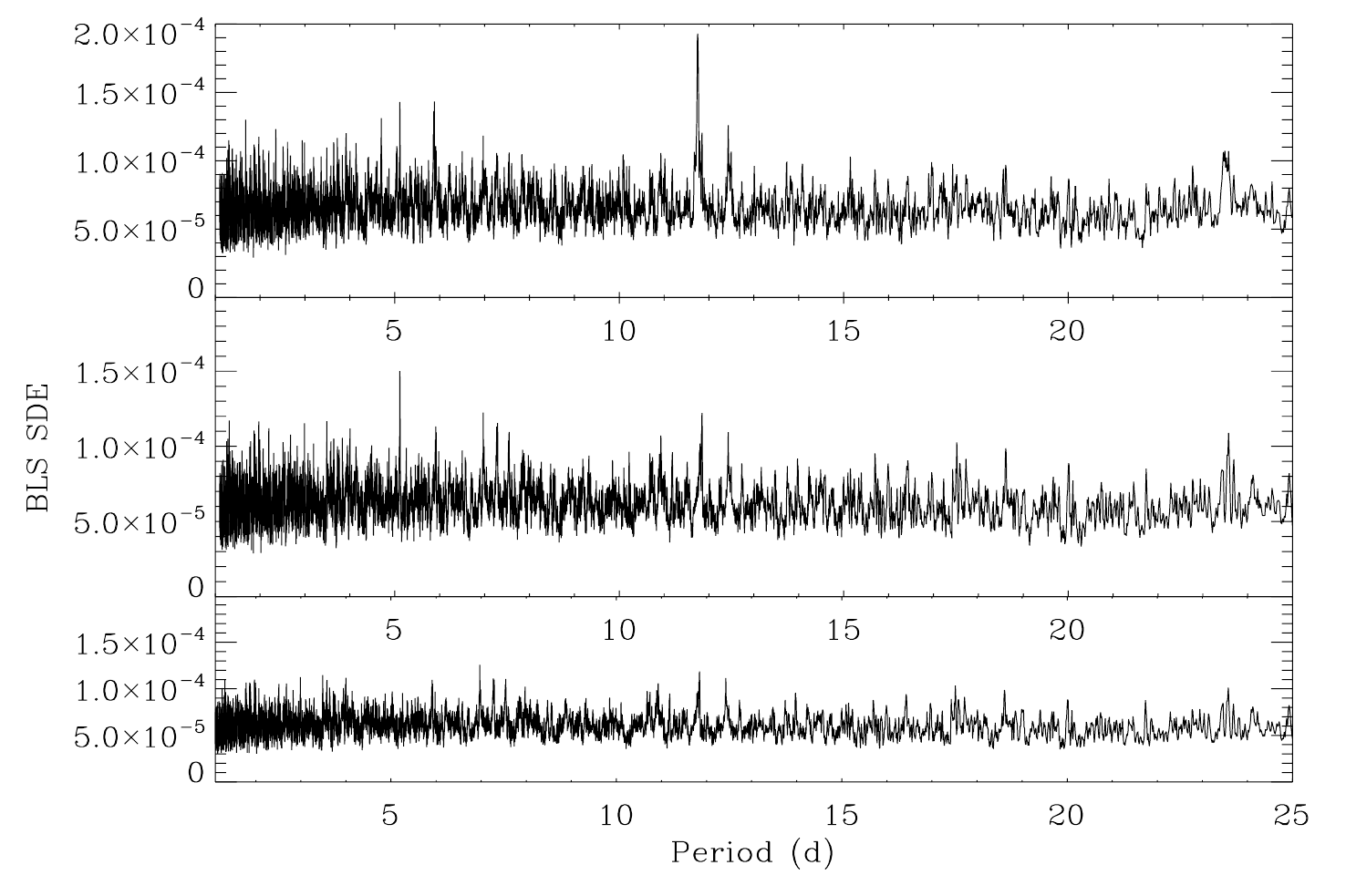}
\caption{SDE of the BLS search of the original normalized curve, showing the peak due to \targetc\ (top); after removal of the transits of \targetc, showing the peak due to \targetb\ (middle), and after removal of $b$ and $c$, showing no significant peaks (bottom).}
\label{fig_bls}
\end{figure}

\begin{figure}[h]
\centering
\includegraphics[width=6.5cm]{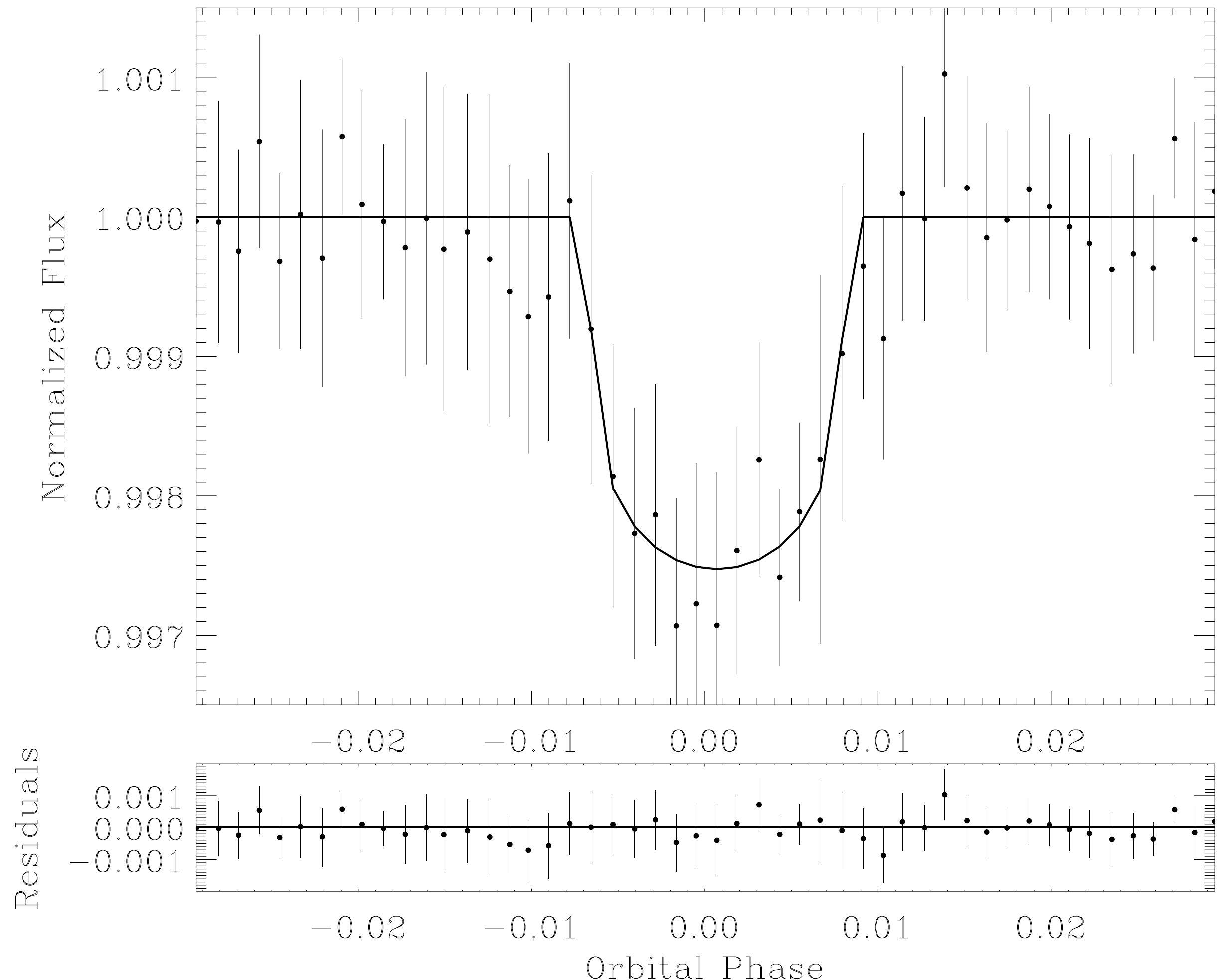}
\caption{Phase folded transit of the 11.75~d period \targetc.}
\label{fig_transit_c}
\end{figure}

\begin{figure}[h]
\centering
\includegraphics[width=6.5cm]{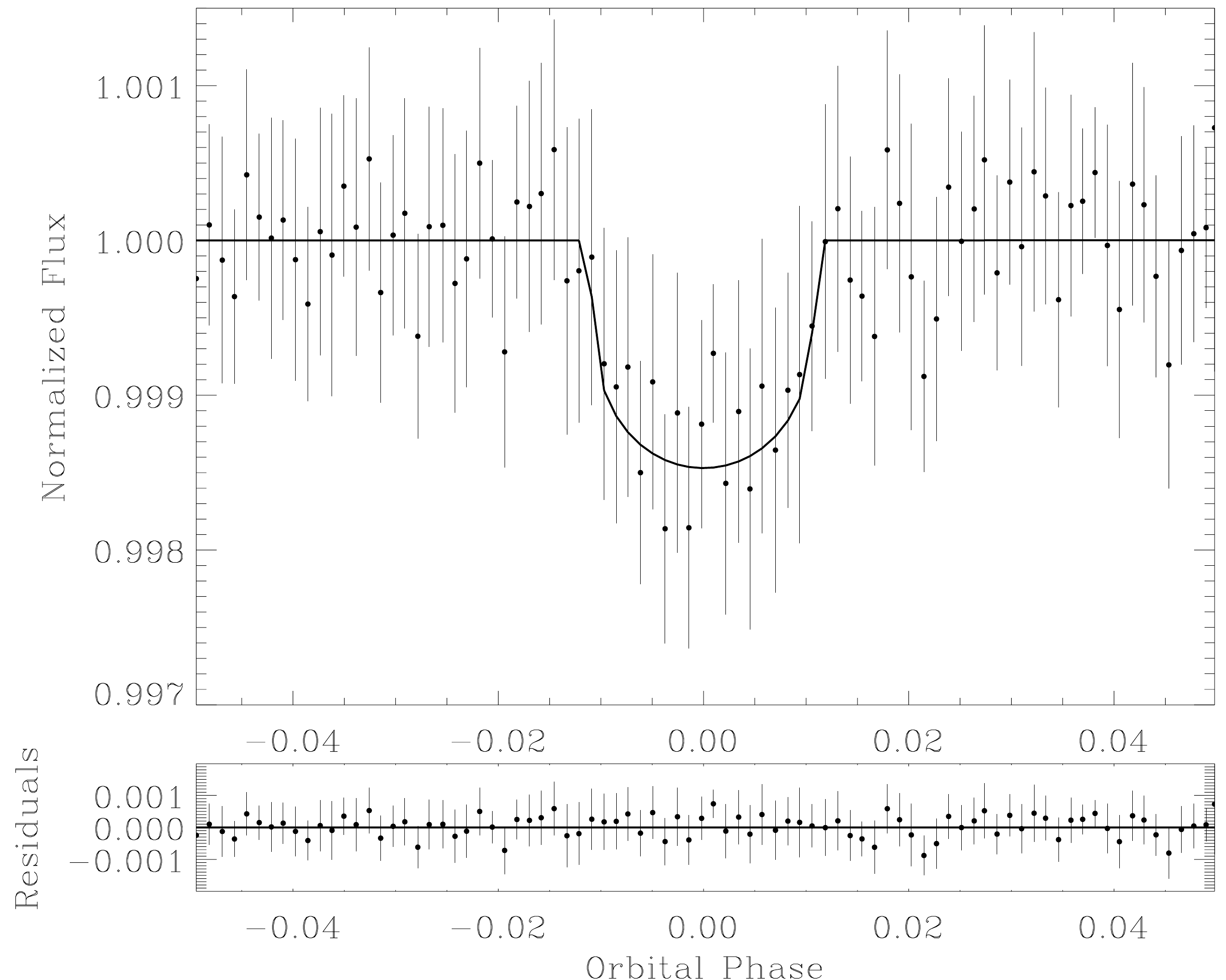}
\caption{Phase folded transit of the 5.11~d period \targetb.}
\label{fig_transit_b}
\end{figure}

A box-fitting least squares (BLS) search \citep{kovacs} shows a peak in the signal detection efficiency (SDE) with an amplitude 13.7 times greater than the standard deviation and no other peaks of comparable amplitude (Fig.~\ref{fig_bls}). Its period of 11.75~d corresponds to the period of the transit identified by the alarm mode, and labelled as \targetc. This transit was modeled with the \cite{gimenez2006} formalism, as done for previous \corot\ discoveries. As in the case of \corot\--7b \citep{leger2009}, the shallow depth of the transits prevents a precise determination of the mean stellar density of \target\, and we rely on the radius estimation from the spectroscopic analysis described in Sect.~\ref{sec: spec_ana}. The fit parameters were the transit center, the radii ratio $k$, and the inclination. The limb darkening coefficients \footnote{Described by a quadratic law on the form $I(\mu)=I(1)[1-u_a(1-\mu)-u_b(1-\mu)^2]$} were estimated from \cite{sing2010} and an error bar of 0.04 was attributed to both $u_a$ and $u_b$. The {\sc amoeba} simplex minimization method \citep{press1992} was used to find the minimum \chisq\ solution, and a residual permutation (prayer-bead) with a random variation of the limb darkening coefficients according to the values given before was used to estimate the uncertainties of each parameter. The eccentricity was fixed to zero. The phase folded light curve and the best fit model are plotted in Fig.~\ref{fig_transit_c}. Once the best fit model was found, the transits of \targetc\ were removed from the light curve. A new BLS search on the residuals shows a peak at 5.11~days with a shallower depth (second panel of Fig.~\ref{fig_bls}) and a peak with an amplitude 11.8 times larger than the standard deviation. This was fit using the same formalism as above. The phase-folded and best fit curves are plotted in Fig~\ref{fig_transit_b} and the candidate was labelled as \targetb\ . As a final step, and to minimize the possible contribution of the transits of \targetb\ to the phase-folded curve of \targetc, we removed the best fit solution of \targetb\ from the light curve and repeated the process to estimate the final values for the parameters of \targetc\ that are given in Table~\ref{starplanet_param_table}.

\section{Ground-based photometry}
\label{sec: groun_phot}

Due to the relatively large point spread function (PSF) of \corot\ that allows one to get color information on the bright targets, there might be several stars inside the PSF that will either cause slightly different parameters obtained from the transit fit if not taken into account, or in the worst case that could have deep eclipses which are diluted by the brightest star in the PSF. Thus, a photometric follow-up with telescopes that provide a narrower PSF is performed in the \corot\ framework. The instruments and techniques are detailed in \cite{deeg2009}, in this section we briefly describe the observations carried out on \target. 

\begin{figure}[h]
\centering
\includegraphics[width=6.5cm]{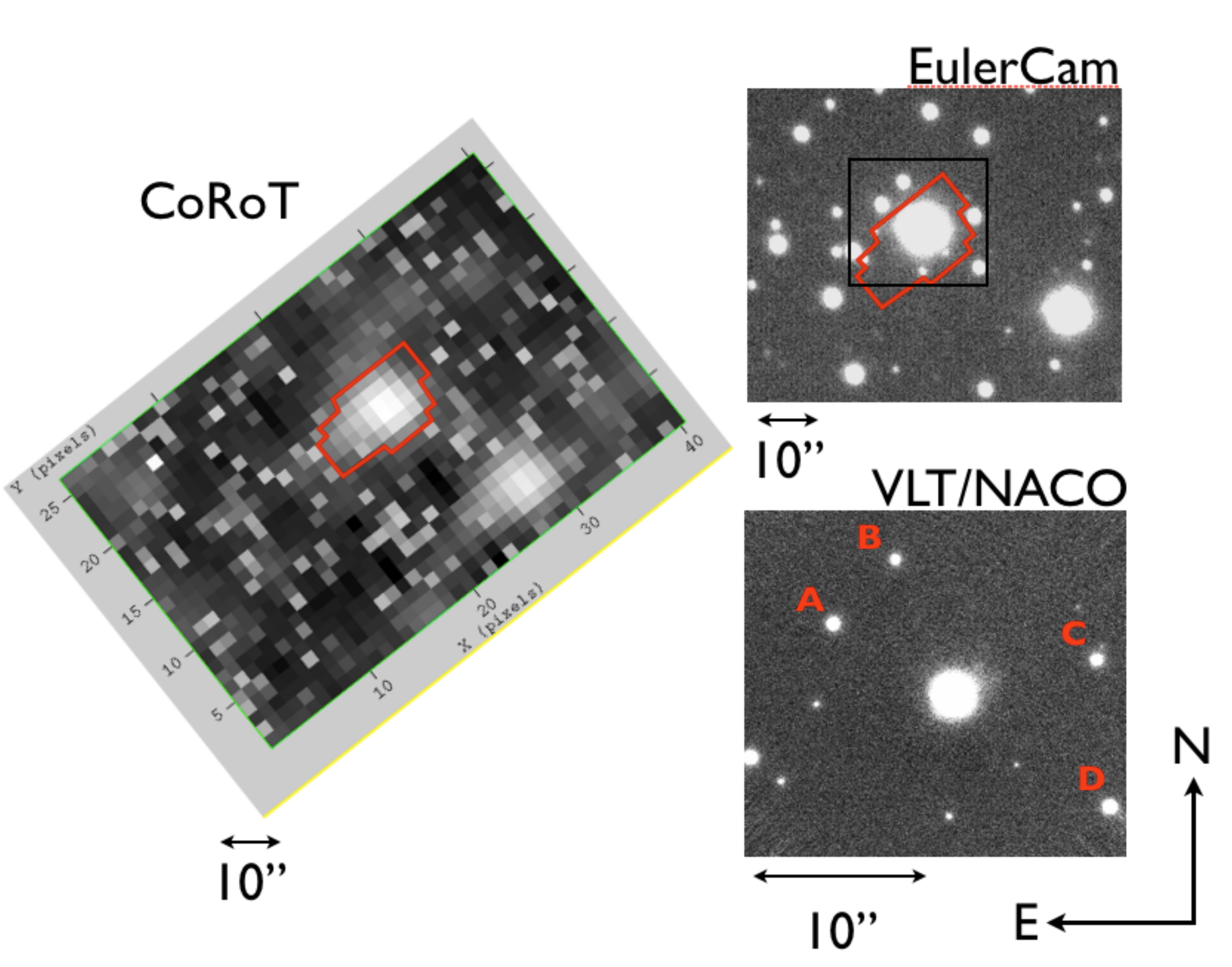}
\caption{Region around \target, from the image of CoRoT taken at the beginning of the observations (left), imaged with the 1.2~m Euler telescope (top right), and with the VLT/NACO (bottom right). The red lines delimit the aperture used to extract the photometry.}
\label{fig_fc}
\end{figure}

The BEST-II telescope \citep{kabath2009} performed pre-characterization observations of the $LRa02$ field between November 2007 and February 2008, and none of the stars close to the target detected with BEST-II showed any variation with \corot\ ephemeris. The PSF of these observations has a typical $FWHM$ of about 6\arcsec, thus higher angular resolution than the \corot\ dispersed PSF of 20-30\arcsec. The transit itself is too shallow to be detected with this instrument. The star at about 28\arcsec\ SW in Fig.~\ref{fig_fc} was observed as LRa02-E1-5235 by \corot\ , and it does not exhibit eclipses or transits at the ephemeris of \target\ b nor~c.

We obtained higher resolution images on both the 82~cm IAC-80 and the  2~m Faulkes Telescope North (FTN) telescopes. These observations revealed at least four faint stars at distances of about 10\arcsec\ from the target, as seen in Fig.~\ref{fig_fc}, and labelled as A, B, C, and D. Only a small amount of the light from these targets is entering into the \corot\ photometric aperture. If we consider the worst -unrealistic- case in which all the flux from these stars is included into the aperture, they would have to show eclipses of at least 30\% in depth to match the depth of the transits of \targetc and at least 17\% to match that of \targetb. The required depth of the eclipses of each star is summarized in Table~\ref{tab:contam}.

\begin{table}
\caption{Required depth of the eclipses in the contaminants A, B, C, D of Fig~\ref{fig_fc}.}
\begin{minipage}[t]{7.0cm} 
\begin{tabular}{lcccc}
\hline 
  & A & B & C & D \\
\hline
Relative R flux (target $=$1) & 0.0075& 0.0057& 0.0080& 0.0058 \\
\hline
To match depth of \targetb\ & 18$\%$ & 25\%& 17.5\%& 24\%\\
To match depth of \targetc\ & 35\%& 46\% & 32\%& 45\% \\
\hline
\vspace{-0.5cm}
\end{tabular}
\end{minipage}
\label{tab:contam}
\end{table}

To check for this worst case scenario and in an effort to try to refine the ephemeris of \targetc, we observed the target in two sequences using the 1.2~m Euler telescope, on January 2nd, 2010, from 00:41 to 09:37 UT, and with the new EulerCam detector on November 12th, 2010, from 5:24 to 08:15 UT. None of the four stars showed variations larger than 20\% for the first sequence, and 10\% for the second one. While the second sequence only covers from slightly after the middle of the predicted transit to the 1-$\sigma$ upper limit of the fourth contact (using the ephemeris from Table~\ref{starplanet_param_table}), the January 2010 observations should have detected an ingress or egress of any of the four stars considering the error in the ephemeris. Since this was not the case and, furthermore, since none of the four stars are fully included into the \corot\ aperture, we conclude that the transits of \targetc\ take place on the target. Three new stars that were closer to the target were detected in the Euler images, but they are too faint to explain the transits of \target\ b or c. 

We checked for eclipses on the four stars during one expected transit of \targetb\ on December 29th, 2011, centered at 06:44 UT using the ephemeris of Table~\ref{starplanet_param_table}. We took images in four different 20-min observing windows centered at about 02:15, 04:15, 05:15 and 06:45 UT, also with the EulerCam detector, and built difference images with a similar observing sequence taken a few days after. None of the images reveals a change in flux of any of the four stars of more than 18\%. Since the 1-$\sigma$ error in the ephemeris at this epoch was 2h50m and the transit duration about 2h51m, these observations discard background eclipsing binaries with the ephemeris of \targetb\ only partially, covering eclipses centered from about -2$\sigma$ to +0.3$\sigma$ from the ephemeris. Further observations are needed to extend this coverage to eclipses happening later than expected.

\begin{figure}[h]
\centering
\includegraphics[width=6.5cm]{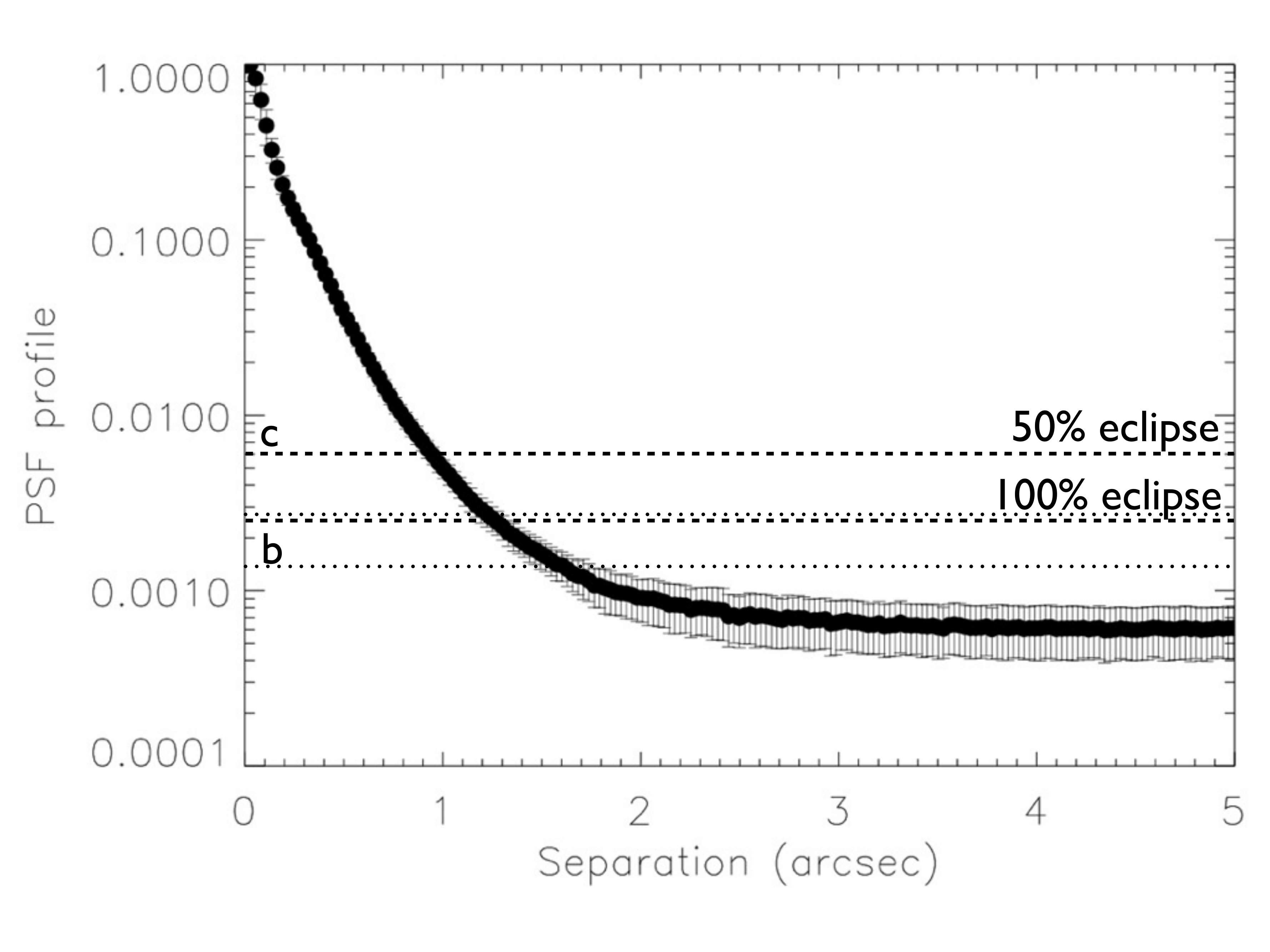}
\caption{Normalised PSF profile around \target, from a VLT/NACO combined image. The horizontal lines delimit the flux level that a contaminant included in the \corot\ 's PSF should reach in order to produce the observed diluted transit depth. Two lines are plotted, for an unrealistic contaminant completely disappearing during eclipse (100\% eclipse line), and a 50\% eclipse. Dashed line is for \targetc, while dotted line is for \targetb. Realistic potential background eclipsing binaries should lie closer than 1.3\arcsec and 0.9\arcsec to the main star to explain the transit depths of \targetb \ and c, respectively.}
\label{fig_naco}
\end{figure}

Finally, we explored the possibility of having other stars closer to the target that lie in the glare of \target\ and that would have not been detected in the other photometric observations. We used the instrument NACO at the VLT to obtain a deep image in $J$ filter with high angular resolution \citep{guenther13}. This filter was selected in order to detect faint background stars that have red colors due to the extinction.The combined image is shown in Fig.~\ref{fig_fc}, where the faint stars that were distinguishable in the Euler images now appear well resolved. A normalized PSF profile of \target\ is shown in Fig.~\ref{fig_naco}, together with the flux limits that a star should reach in order to reproduce the transits of \target\ b and c, for the conservative cases of a 100\% eclipse and a 50\% one. To remain undetected in this high resolution image, potential eclipsing binaries should lie closer than 1.3\arcsec and 0.9\arcsec to the target to explain the targets of \target\ b and c, respectively. \footnote{These numbers are obtained under the assumption that there is no significant color difference between the stars observed in the CoRoT bandpass vs. the NACO $J$-band image. A background star redder than the target would require deeper eclipses in the NACO image to explain the observed depths in CoRoT. This would shift the horizontal lines of Fig.~\ref{fig_naco} upward, resulting in more stringent limits to the distance of the stars to remain unresolved in the NACO image.} 

\section{Radial velocities}
\label{sec: rv}

\begin{figure}[h]
\centering
\includegraphics[width=9cm]{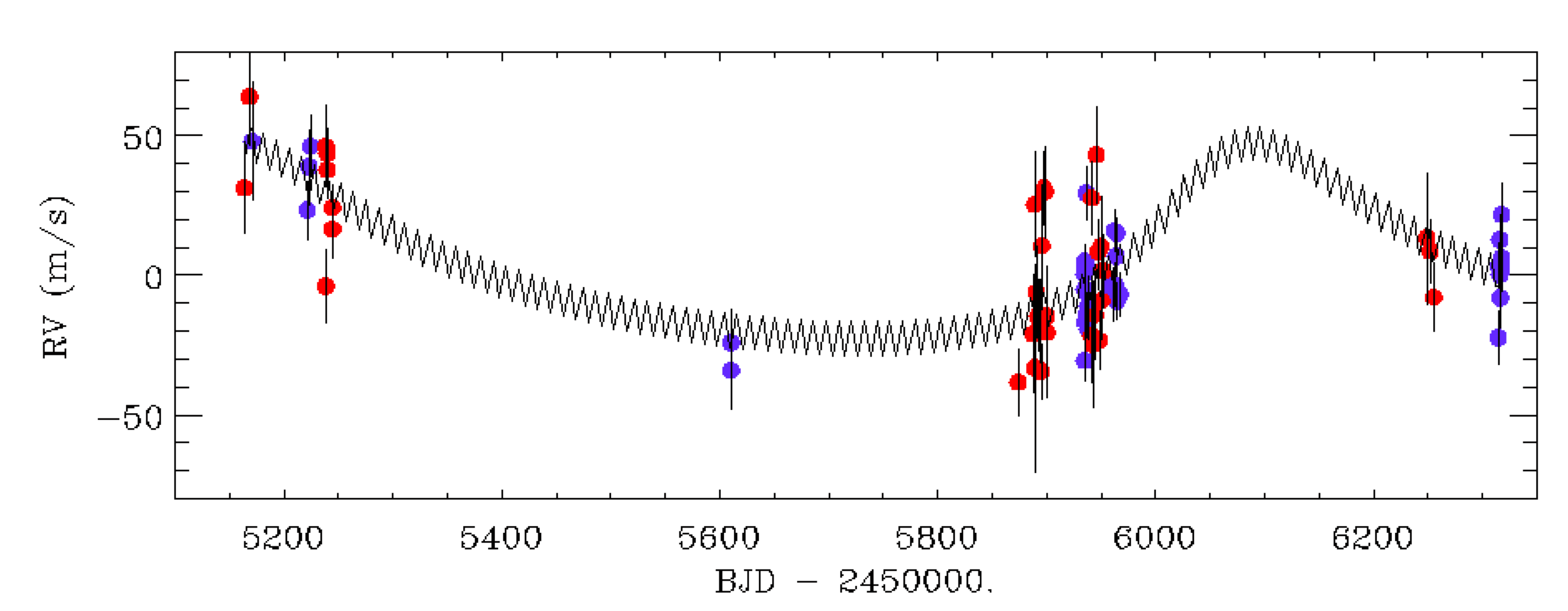}
\caption{Radial velocities, showing a long-term trend compatible with a 1.5~\Mjup\ object with a period of $~$940~d. Blue points for HIRES data, red points for HARPS. More data are needed, especially at the phases of maximum RVs.}
\label{fig_RV}
\end{figure}

Two of the spectrographs that have shown to reach the precision required to discover super-Earth like planets were used in an effort to measure the masses of the planets in the system (see \citealt{santerne2011} for details on the \corot\ radial velocity follow up strategy). 37 spectra were taken with the HIRES/Keck1 spectrograph \citep{vogt94} as part of NASA's key science project in support of the \corot\ mission, and 34 measurements were performed with the HARPS/ESO-3.6 spectrograph \citep{mayor2003}. These measurements  are summarized in Table~\ref{rv}, and plotted in Figure~\ref{fig_RV}. The data show a long term modulation that is compatible with the presence of a longer period object in the system, with a period of $\sim$940~d, a K$\sim$35~m\,s$^{-1}$ and an eccentricity close to 0.4. It would correspond to an object with a minimum mass of $\sim$1.5~\Mjup\ and a semi-major  axis of 1.8~A.U. The obtained parameters for this candidate planet rely strongly on the two single data points taken in 2010, which sample the phases of minimum observed radial velocity. Further observations are thus encouraged to confirm and better constrain the orbital parameters of this candidate signal, especially around the phases of the next predicted maximum radial velocity, which should happen around January, 2015. 

\begin{figure}[h]
\centering
\includegraphics[width=8.5cm]{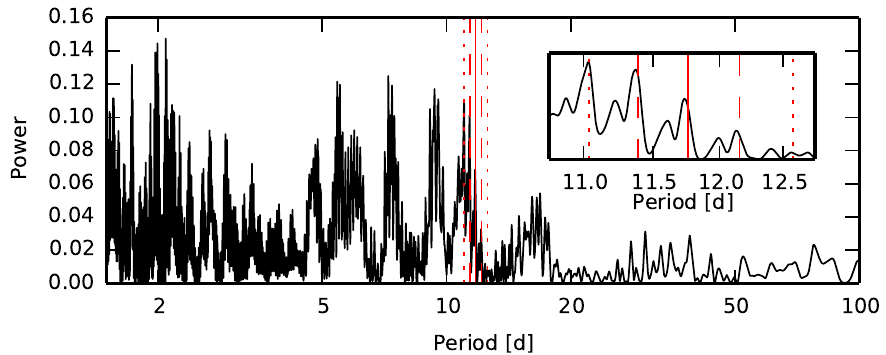}
\caption{Periodogram of the combined HARPS + HIRES data set, after the removal of the $\sim$940~d signal. A low-significance excess of power around the period of \targetc\ (11.759~d) is detected, and zoomed in the inset plot. The vertical lines show the positions of the photometrically determined period, and the dashed and dotted lines the locations of the harmonics due to the observational window.}
\label{fig_periodRV}
\end{figure}

The periodogram of the combined HARPS + HIRES data set, after the removal of the long period signal, is shown in Fig.~\ref{fig_periodRV}. Several peaks are identified at around 27.3~d, 13.5~d, 9~d and 6.6~d, that could correspond to the rotational period of the star and its harmonics (P$_{\rm rot}$/2, P$_{\rm rot}$/3, P$_{\rm rot}$/4). A rotational period of 27.3~d is compatible with the value of the \vsini\ reported below assuming $i_\star$ close to 90$^{\circ}$ and with the autocorrelation of the photometric curve. 

The region close to the photometrically determined period of \targetc\ shows increased power, taking the observational RV window function into account. In fact, the peak with most power is consistent with the second harmonics of the window function, and it is found at $\sim$11.038~d. Since the orbital period is well determined from the photometry, we evaluated the false alarm probability of this signal by constructing sets of mock data shuffling the RV data points, keeping the same time stamps. We computed the integrated power of the periodogram in the regions between the period of \targetc\ and three times the error listed in Table~\ref{starplanet_param_table} and in each of the harmonics due to the window function (which shows two peaks in frequency at 0.00277 and 0.005535~c/d, corresponding to the annual and seasonal observing windows). The false alarm probability was estimated by counting the number of times the integrated power in these regions was higher than in the data, resulting in a value of 11\%.

Under the assumption that the RV signal from the two transiting candidates is present in the data, we first estimated the respective $K$-amplitudes of \targetb\ and \targetc\ by performing a two-sine component fit to the data after removing the long-term trend. The periods and phases were fixed to the transit ephemerides. The orbital solution was calculated using the general nonlinear least squares fitting program GaussFit \citep{jefferys88}. This resulted in $K_b$ = 0.80 $\pm$ 3.0 m\,s$^{-1}$ and $K_c$ = 5.2  $\pm$ 3.2 m\,s$^{-1}$. This suggests that any short-term variations in the RV measurements are dominated by \targetc.

We then fitted the RV measurements using a variation of the floating offset fitting technique employed in detecting CoRoT-7b \citep{hatzes11}. The RV data were divided into eleven subset data, four for Keck and seven for HARPS. The time span of each data subset ranged from two to seven days, with a median value of five days. On several nights there were three Keck RV measurements and for the fit we used the median value. An orbital fit was made keeping the period and phase fixed to transit values, but allowing the $K$-amplitude and zero-point offsets of the subset data to vary. 

The basis for this method is that during the short time interval spanning each subset the variations due to the long-term trend or stellar variations from rotational modulation is relatively constant. Although the time span of the subset data is a significant fraction of the rotational period (0.07 to 0.17) the star has a relatively low level of activity, so our assumption seems reasonable. This technique has the advantage of removing the long-term trend without a functional fit. It also provides a natural way of combining the HARPS and Keck data, each of which have their own velocity zero-point.

The floating offset fit to the data using the period of \targetc\ resulted in  $K_c$ = 8.31 $\pm$ 2.86 m\,s$^{-1}$. Applying the method using the period of \targetb\ resulted in $K_b$ = 0.88 $\pm$ 2.46 m\,s$^{-1}$. We then tried fitting the residual velocities using the period of \targetb, but after first subtracting the contribution of the derived orbit of \targetc. This resulted in $K_b$ = 0.0 $\pm$ 2.1  m\,s$^{-1}$. The floating offset method confirmed the results found by the two-component sine fitting, namely that \targetc\ is marginally detected (2.9$\sigma$) while \targetb\ is undetected.

Fits  using the individual RV data sets gave consistent results. A fit using only the Keck data resulted in $K_c$ = 6.64 $\pm$ 3.60 m\,s$^{-1}$. The fit to only the HARPS data resulted in $K_c$ = 10.76 $\pm$ 4.62 m\,s$^{-1}$. 

\begin{figure}[h]
\centering
\includegraphics[width=8.5cm]{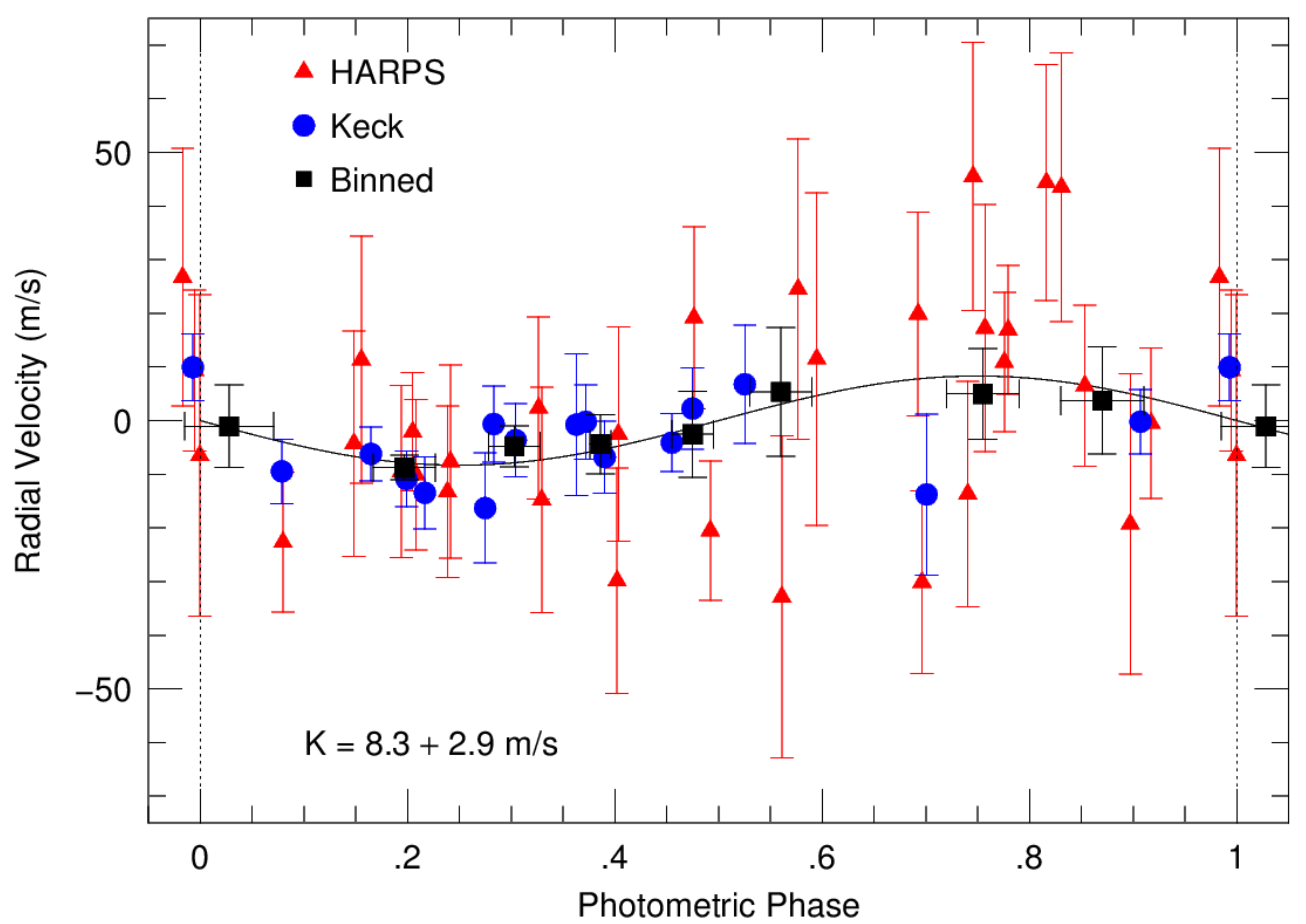}
\caption{Radial velocity measurements folded at the period of \targetc, and best fit orbital solution with the parameters given in the text. Blue dots are for HIRES, red triangles for HARPS, and black squares data binned for displaying purposes.}
\label{fig_rv2}
\end{figure}

Fig.~\ref{fig_rv2} shows the RV variations phased to the transit ephemeris of \targetc\ after removing the calculate zero-point offsets from the respective subset data. To better see the RV variations we also showed binned values.

The obtained semi-amplitudes correspond to a 1-$\sigma$ upper limit for the mass of \targetb\ of 5.7~\Mearth\ and a mass of 28$^{+11}_{-11}$\Mearth\ for \targetc\, taking the host star mass estimation described in the next section into account.

\section{Spectral analysis}
\label{sec: spec_ana}

Seven HARPS spectra of \target that were not contaminated by the moon light were co-added once set in the rest frame of the star. This provides us a spectrum with a S/N-ratio of 88 per element of resolution. We determined the photospheric parameters of the host star using the same methodology as in past \corot\ papers: the spectroscopic analysis was done independently using the semi-automated package {\sc vwa} (\citealt{bruntt2010}, and references therein) and {\sc sme} \citep{valenti1996}. The parameters obtained are listed in Table~\ref{starplanet_param_table}. The mass and radius of the star were then estimated using the T$_{\rm{eff}}$, metallicity and log$g$ derived from the spectroscopic analysis, to fit the evolutionary tracks of {\sc starevol} (Palacios, priv. com., see \citealt{turck10,lagarde12} for a description of the code). It yields a mass of $M_\ast$ = 0.91$^{+0.08}_{-0.07}$~\Msun\, and $R_\ast$= 0.86$^{+0.03}_{-0.04}$~\Rsun\, for the star. The errors are formal statistic errors, which are known to provide unrealistically low values due to, among other things, the intrinsic errors of the stellar models. We thus adopted in Table~\ref{starplanet_param_table} error values of 10\% for both mass and radius of the star, to account for these extra errors.

We used the derived photospheric parameters and the available photometry to fit and scale a model spectrum, in order to determine the interstellar extinction and the distance to the system. The spectral energy distribution (SED) nicely follows the scaled model, as shown in Fig.B.3 of \cite{guenther13}, and shows no evidence of any mid-infrared excess. 

\begin{figure}[h]
\centering
\includegraphics[width=8cm]{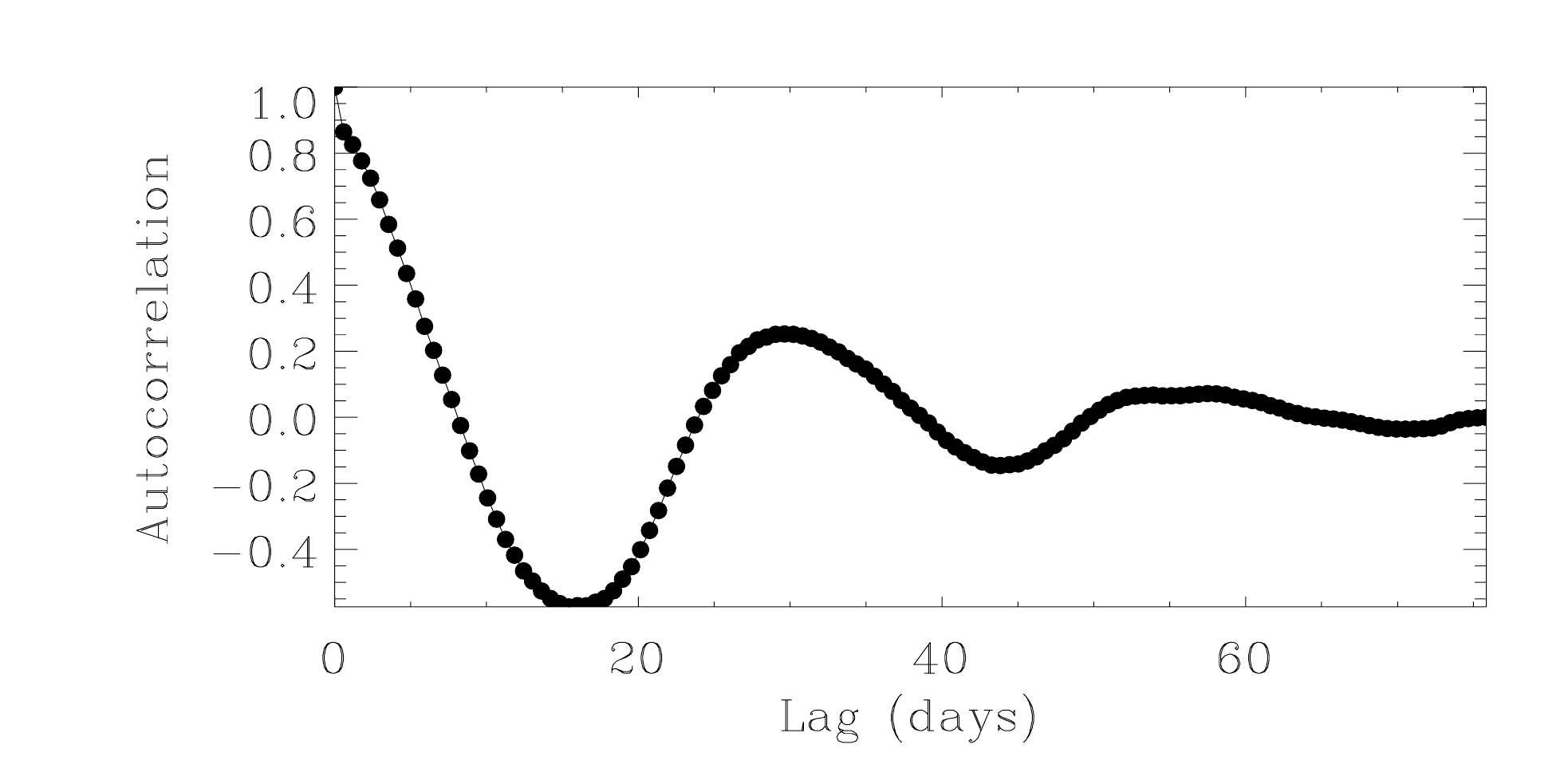}
\caption{Autocorrelation of the normalized light curve.}
\label{fig_corr}
\end{figure}

The projected stellar rotation velocity was determined by fitting the profile of several well-isolated spectral metal lines. We found \vsini = 2.0$^{+1.0}_{-1.5}$\kms. The autocorrelation of the light curve shows a peak at about 29~days (Fig.~\ref{fig_corr}), which is consistent with a rotation with the \vsini\, found above. As in previous cases, we found models that reproduce the spectroscopic parameters corresponding to a pre-main sequence evolutionary phase. Due to the \vsini\, and the estimated rotation period, we did not consider these solutions, because the star should be rotating at higher velocity. The age of the star in the main sequence solutions is around 11~Gyr.

In order to search for late-type stars close to \target, whose brightness difference is much smaller in the infrared than in the optical, we obtained a high
resolution infrared spectrum using the CRIRES spectrograph
which is mounted on the VLT-UT1 (Antu), as part of a program for follow up of CoRoT candidates. The strategy, instrumental configuration and results for several candidates, including \target, is described in \cite{guenther13} (the case of \target\ is described in its Appendix B.6).  These observations exclude non-resolved physically bound companions with spectral types earlier than M2.5V. 

\section{Blend simulations}
\label{sec: simul}

Without a very significant detection of the radial velocities, we need to carefully check different blend scenarios that might explain all the observations described in the previous sections. In the next paragraphs, we describe the technique we used to simulate and evaluate the probabilities of the transit signals to be due to different types of stellar systems involving eclipsing binaries or stars with bigger transiting planets. The method is quite similar and strongly inspired by the {\sc blender} software described in \cite{torres2011} and \cite{fressin2011}. We apply it to systems with either four or five bodies, because we consider the two periodic transit signals real, and there is no stellar system that can remain stable with such short periods. The simulations of systems with four bodies account for those configurations in which there is one transiting planet around one star, and an eclipsing binary that can be physically associated or in the line of sight. Simulations with five bodies are for systems where the star 1 (the one detected in the spectra) is blended with two stars, each in an eclipsing system or containing a transiting planet, also at a wide variety of distances. 

As a starting point, we used the cross-correlation function (CCF) of the HARPS spectra to put flux limits to other stars that are included in the spectrograph's fiber. The main peak of the CCF was subtracted from the individual exposures, and fake CCFs with different widths were inserted at different velocities. We quantify a 3-$\sigma$ level for non-detection of CCFs corresponding to flux ratios of 0.01 of the main star. This condition was included in the simulations described below.

We constructed light curves with the contribution of one of the planets removed according to the results of the fits described in the previous section\footnote{We did not consider the case of mutual transit occultations during the transit. In the time span of the observations there is a less than one hour overlap of the transit events centered at about 3285 in the Fig.~\ref{fig_lc}. Unfortunately, the photometric precision does not allow checking if there are mutual occultations in that event.}. For each of these curves, we built a phase folded light curve with the period of the remaining transit. The goal of having a whole orbital phase curve for each of the transiting objects is to check for potential secondary eclipses at any of the orbital phases, since eccentric false positives will also be checked. To construct these curves, we normalized each of the planetary orbits with a 7th order polynomial, thus removing most of the variations due to the stellar activity or uncorrected instrumental effects. We checked that these low order polynomials would not be able to remove the short (compared to the orbit) secondary eclipses. Finally we combined all the orbits into a normalized phase curve binned in 0.0012 in phase (total of 834 points). To check for systems whose period is twice the one of the potential planet, which correspond to stellar systems having similar primary and secondary eclipses, we also built curves folded at twice the period. In total, four light curves were built this way, two for each of the transiting signals to be studied.

These light curves will be compared to many different models of four or five body systems, as described before. In the five body scheme, we will call star $1$ the one whose parameters were extracted in the spectroscopic analysis of Section~\ref{sec: spec_ana}, stars $2$ and $3$ are the components of one of the eclipsing systems, and stars $4$ and $5$ the components of the other. In the four body scheme, the only eclipsing binary is composed of stars labelled $3$ and $4$.

For each of the models, the stellar parameters of star $1$ are fixed to the results of the spectral analysis. We interpolate the ATLAS9 model atmospheres of \cite{castelli2004} to the values of the T$_{\rm{eff}}$, log $g$ and metallicity, scale it with the stellar radius and its distance, apply an extinction law of \cite{fitzpatrick1999} to the $E(B-V)$ of the galactic interstellar extinction model of \cite{amores2005}, and finally apply the spectral response function of  \corot\ given in \cite{Auvergne2008} to get the observed flux of star $1$.    

\subsection{Four body simulations}

The star $1$ is orbited by one planet ($2$), and we simulated eclipsing binaries (systems of star $3$ and $4$, where $4$ is also allowed to be a planet as a particular case).  We explored systems where the period of the planet $2$ is that of 5.11~days, and systems where it was 11.76~d. In each case, the system $3$-$4$ will be assigned the other period, or twice this value. This is done to account for the possibility of star $3$ and $4$ being similar, and thus having no significant differences between primary and secondary eclipses. We extracted one value for the age and mass of the star $3$-star $4$ eclipsing system in the model from an uniform random distribution. We also chose a value for the eccentricity and the longitude of the periastron randomly from an uniform distribution, and a random distance for the eclipsing system. This distance is allowed to vary from 0 to 10~kpc for the observations towards the \emph{anticenter} direction. Once these random values have been generated for stars $3$ and $4$, we interpolate the isochrones of \cite{marigo2008} \cite{girardi2010}\footnote{http://stev.oapd.inaf.it/cgi-bin/cmd}, assuming solar metallicity, to get the values of the T$_{\rm{eff}}$ and log $g$. Then the ATLAS9 model atmospheres are interpolated to these values and scaled with the stellar radius and distance coming from the isochrones. We apply the same extinction law, model and \corot\ response function to these spectra, and obtain the observed fluxes of stars $3$ and $4$. The semi-major axis is calculated from the masses and the period of stars $3$ and $4$ and the orbital inclination is estimated from the primary eclipse duration with the associated error bars (measured in the light curve), using Eq.~7 in \cite{tingley2005}. 
We can then compute the model light curve of the star $1$-planet $2$ and of the system star $3$-star$4$, using the {\sc jktebop} code (\citealt{south2004a,south2004b}, based on {\sc ebop} \citealt{popper1981,etzel1981}) and the limb-darkening coefficients of \cite{sing2010}, dilute the system using the observed flux of star $1$, and calculate the \chisq\ with respect to the \target\ phase light curves. This process is repeated for many (6$\cdot10^7$) different random realizations of the parameters of the star $3$ and $4$, and of planet $2$. In order to speed-up the process, several conservative conditions are plugged in the system in order not to compute unnecessarily the eclipsing binary models: for instance, if there are no eclipses that will match the observed duration, or if the undiluted eclipse depth will be in any case smaller than the observed transit depth. We also impose the condition to have the star $1$ brighter than the total of star $3$ + star $4$. 
In this case, our models have 13 free parameters, five for the planet around star $1$ (mass, radius, inclination, eccentricity and omega), and eight for the star $3$-star $4$ system (age, initial masses of $3$ and $4$, distance, inclination, eccentricity, omega, and possibility of having 2x period). In the case of the body $4$ being a planet the number of free parameters remains unchanged (age, initial mass of $3$, radius and mass of the planet, distance, inclination, eccentricity, omega). 

\subsection{Five body simulations}

We simulated two eclipsing systems, namely star $2$-star $3$, and star $4$-star $5$. In each of the cases the period is allowed to be one of the periodic signals found in the light curve, or twice this value. Solutions where the system $3$ or $5$ is a planet were also included. The total number of free parameters is 16, i.e., 8 for each of the systems, the same as in the case of the star $3$-star $4$ scenario of the four body simulations.

\subsection{F-tests}

To compare the blend and multiple transiting planetary system scenarios we used the $F$-test, that is based on the ratio of two reduced \chisq\,. The $\chi^2$ maps were built in the parameter space mass of star 2 (or 3, or 4, depending on the case) vs distance modulus ($\Delta \delta=5\;\log_{10}\left(\frac{D_{blend}}{D_{star 1}}\right)$) for each scenario. In each box of the two-dimensional grid, we took the minimum $\chi^2$ of all the models inside that box, and we computed $F=\chi^2_{reduced,blend}/\chi^2_{reduced,planetary system}$. 

\begin{figure*}
\centering
\includegraphics[width=\textwidth]{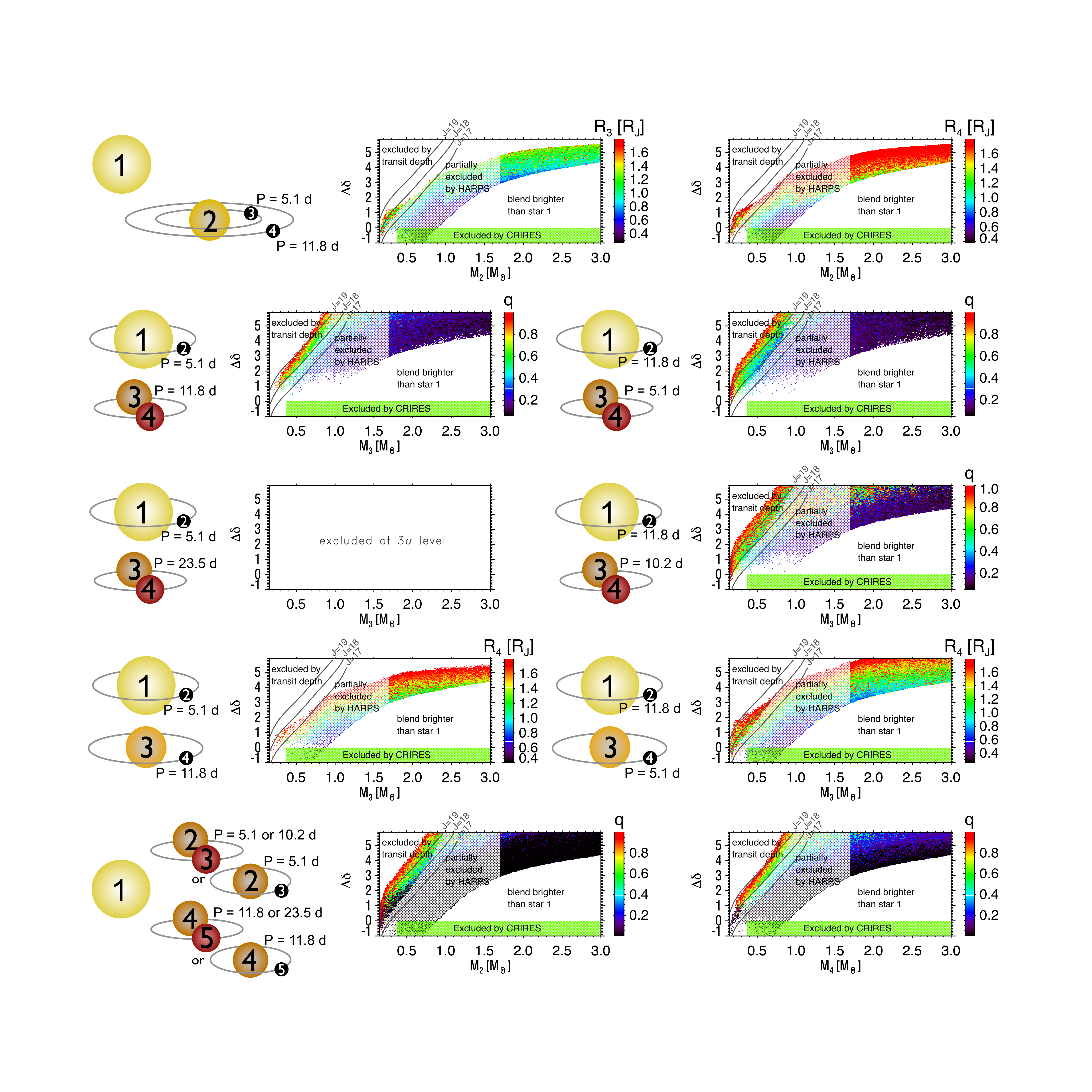}
\caption{Maps showing the regions where the $F$-test resulted in solutions not excluded to a level above 3-$\sigma$ with the multiple transiting system solution (solutions excluded above 3-$\sigma$ appear white in the plot). The different explored configurations are sketched at the left of each plot. The color scale is for the different q ratios of the binary star or for the radii of the planets, and the color intensities are for different levels of exclusion (more intense meaning less excluded). The ordinate axis shows the distance modulus ($\Delta \delta=5\;\log_{10}\left(\frac{D_{blend}}{D_{star 1}}\right)$). Bound, or closer to the observer, second stars are excluded with CRIRES spectra for spectral types later than M2.5V \citep{guenther13}. Regions labelled as ``partially excluded by HARPS" are those in which a second CCF would have been detected in the HARPS data. In the case of rapidly rotating stars, this second CCF might have been unnoticed. In every case, second stars with masses above 1.7\Msun\ have been assumed to be undetectable in the CCFs, what explains the vertical cuts at this value. The upper left parts of the plots are typically excluded because the hypothetical system would produce eclipse depths below the observed values. The diagonal solid lines labelled with different $J$ values are plotted to provide a direct comparison with the exclusion values from the NACO $J$-band images given by \cite{guenther13}: a $J$=17 star would have been detected at (and above) 0.18\arcsec from the primary, a $J$=18 at 0.23\arcsec, and a $J$=19 at 0.58\arcsec. }
\label{fig_big_blend}
\end{figure*}

The results for all the different configurations explored are plotted in Fig.~\ref{fig_big_blend}. Since the radial velocity results provide a clearer detection of the \targetc\ signal, we describe in more detail the results of the simulations in which one of the stars has a transiting planet with the period of \targetc\ and explore the possible scenarios that are compatible with the data. There are typically two regions in the maps that are not excluded above a 3-$\sigma$ level.  One corresponds to eclipsing binaries with the primaries being hot and massive stars, with masses higher than 1.7\Msun.  In these cases, a CCF might have not been detected in the radial velocity observations, but they need to lie at large distances in order not to become the brightest member of the system, what would give results that are inconsistent with the spectral analysis. The other set of solutions is for eclipsing binaries with masses of the primary below 0.9\Msun, at a distance in which the diluted eclipses could be mistaken with the transits of a planet, but not so close as to have been detected in the CCF of the spectra.

The five body results are more restrictive, as they require none of the primaries of the two blended eclipsing systems to be detected in the CCFs, and each of the systems needs to lie at a particular physical distance with respect to the brightest object (star $1$).

The first row of Fig.~\ref{fig_big_blend} shows the case of two planets orbiting around star $2$, that after dilution created by star $1$ would appear as small planets. Most of the allowed solutions are for sizes of the planets of $>1$\Rjup. Such a system with two Jupiter-sized planets orbiting that close to their star has not been discovered yet, and multiple transiting systems are known to rarely harbour a transiting giant planet \citep{latham11}.

\section{Discussion}
\label{sec: disc}

\subsection{Probabilities for the different scenarios }
\begin{figure}
\centering
\includegraphics[width=8.5cm]{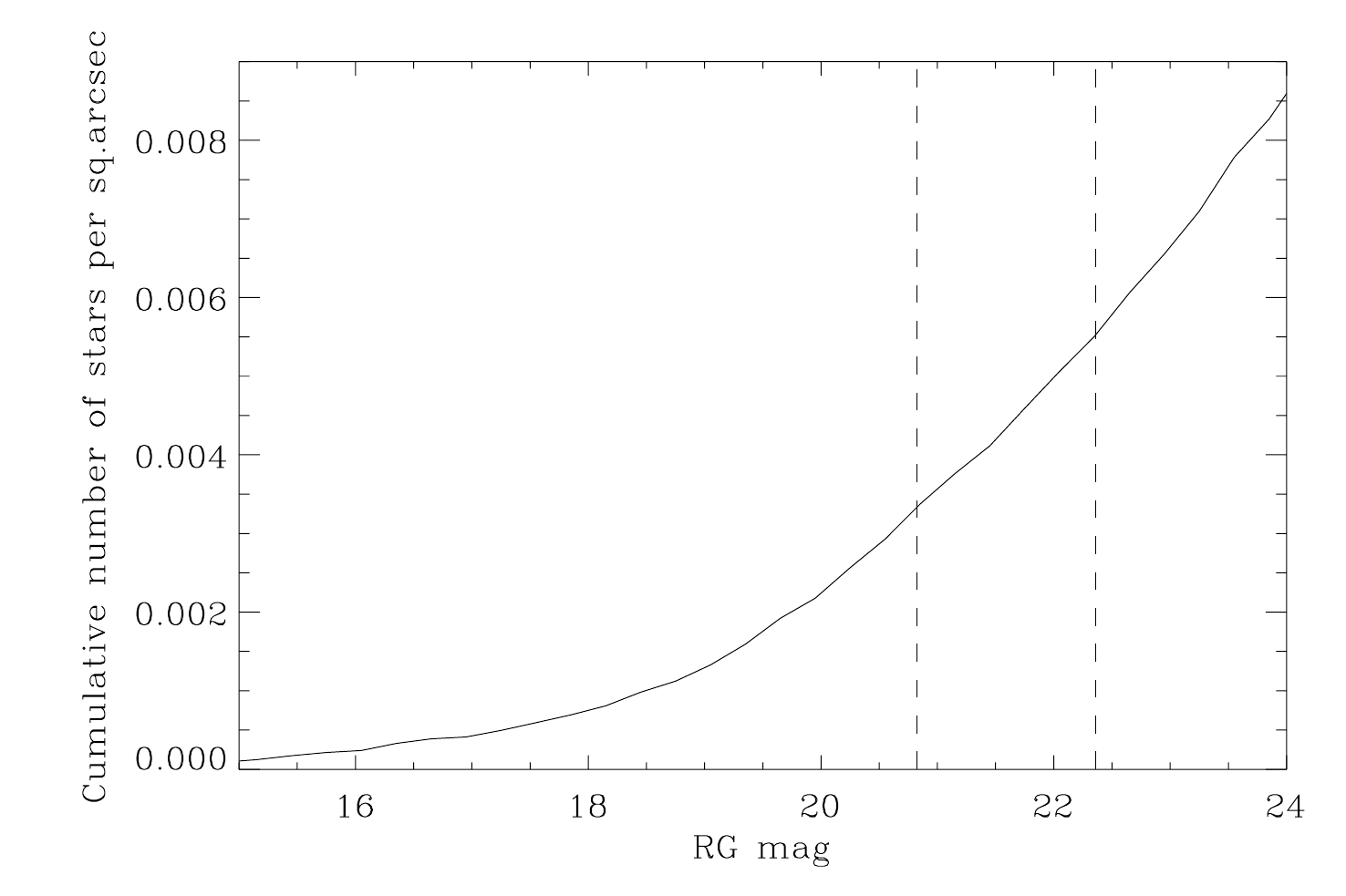}
\caption{Cumulative number of stars per square arcsec, computed from a combined image of the field of \target\ that is complete down to RG$\sim$25. The two vertical lines correspond to the extremes of the horizontal lines in Fig.~\ref{fig_naco}, i.e., the highest magnitude a star could have in order to reproduce the observed transit depth of \targetb\ while completely disappearing during eclipses, and the magnitude of a star with 50\% eclipses needed to explain the transit depth of \targetc. }
\label{fig_cum}
\end{figure}

We used an average image from the second Euler transit sequence in order to estimate the density of stars in the direction of the \target\ observations. The $FWHM$ of the PSF in this image is about 1\arcsec. Fig.~\ref{fig_cum} shows the cumulative number of stars per square arcsec, and the upper magnitude limits that a star can have in order to reproduce the observed depths, neglecting color effects that might move these lines slightly if the colors of the contaminants are different between the CoRoT and the RG bandpasses. From this Fig., we can (conservatively) estimate the odds for a random alignment of a background/foreground star, that would remain undetected in the NACO high resolution images, and whose brightness is enough to explain one of the two transiting signals, to about 1:180 (=1:1/0.0055, corresponding to the crossing of the vertical line on the right of the plot and the solid line). To this, we need to add the fact that the non resolved star should be an eclipsing binary. The fraction of eclipsing binaries found by \corot\ in previous fields (IRa01: 145 out of 9872 sources, 1.47\% \citealt{carpano2009}, LRc01: 158 out of 11408 sources, 1.38\% \citealt{cabrera2009}) is similar in the center and anticenter directions, and comparable to the results of the \kepler\ mission (2165 out of 156453 sources, 1.38\% \citealt{slawson2011}). Out of these, close to 50\% are not ``appropriate" to be confused with a transiting planet, as they belong to the contact or semi-detached systems. Thus, the fraction is on the order of 0.7 -- 0.8\%, and the odds to have a star that is an eclipsing binary non resolved in the NACO image drop to 1:22000 -- 1:26000. 

According to the results of the previous section, the eclipsing binary has also to lie at a particular physical distance from the main target, and have an adequate primary mass and q-ratio in order to reproduce the photometric observations. The constraints imposed by the CCF and the photometry exclude most\footnote{some particular -and low probable- systems with similar spectral type as the target, and the same systemic velocity, or in which one of the stars has a very fast rotation that avoids the detection of the CCF, and are not resolved by the NACO images, can not be excluded. This statement holds for many of the currently known exoplanets.} binaries with a primary with a mass from 0.9 to 1.7~\Msun. Slightly different distances or q-ratios could produce blended transits with depths that would be confused with that of a planet of different size as that of \targetb. Using the distribution of detached eclipsing binaries of \cite{slawson2011} we consider that we can safely exclude up to 50\% of the potential eclipsing binaries because their primaries are in the mass range mostly excluded by the CCF. We estimate the odds then to have a star that is an adequate eclipsing binary, non resolved in the NACO image, and not detected in the CCF of 1:44000 -- 1:52000.

Because the radial velocities of the brightest star do not show variations at the typical levels for stellar companions (km/s) and since stellar eclipsing triple systems with periods of five and eleven days are dynamically not stable, we have to estimate the probability to have two adequate eclipsing binaries inside the \corot\ PSF. With the photometric follow-up we have excluded the case where the \targetc\ eclipses happen in any of the stars that are at about 10\arcsec\ in Fig.~\ref{fig_fc}. But, due to the non complete verification for eclipses with the ephemeris of \targetb\ (excluded in a window from -2$\sigma$ to +0.3$\sigma$ of the predicted ephemeris), we can currently only estimate the probability that one of the five potentially worrying stars is an adequate eclipsing binary as $\sim$5:130 (5 stars -the 4 labelled in Fig.~\ref{fig_fc}, plus an hypothetical star non resolved in the NACO images-, and a 0.7-0.8\% fraction of appropriate eclipsing systems). Thus, the probability to have a blend of two eclipsing binaries and a single target star can be currently estimated as $<$8.8$\cdot$10$^{-7}$ (=1:44000$\times$5:130). If future photometric campaigns are able to definitely reject the possibility of having eclipses in the detected stars at projected distances of about 10\arcsec\ from the target (see Sect.~\ref{sec: groun_phot}) then this probability could drop to $<$5.2$\cdot$10$^{-10}$ (=1:44000$\times$1:44000).

 This is to be compared with the probability of having a genuine Neptune transiting system. Using the results from the \kepler\ mission \citep{borucki2011}\footnote{http://exoplanetarchive.ipac.caltech.edu/ , we used the targets labelled as ``CANDIDATE" in the ``Disposition using Kepler data" column, and the Q1-Q8 catalogue of \citep{burke14}.}, we count 162 planetary candidates with sizes between 3.3 and 5.5 times the size of the Earth, out of the $\sim$190,751 stars surveyed by the mission. Assuming that all these are true planets, we get that a fraction of 0.85$\cdot$10$^{-3}$ of the stars host a planet with a size in the range of the candidates around \target. Alternatively, we can also estimate the fraction of multiple transiting systems with two transiting components using the results of \cite{lissauer14}, as 272 candidates out of the observed stars with \kepler, representing a fraction of 1.4$\cdot$10$^{-3}$. This later number includes also systems with sizes smaller than those detectable in the \target\ light curve, and thus constitutes an upper limit. 
 
Thus, we estimate the five body scenario as $>$966$\times$ less probable than having at least one transiting planet around the main target (=0.85$\cdot$10$^{-3}$ / 8.8$\cdot$10$^{-7}$). 
 
Out of the four body scenarios, in which there is one genuine transiting planet and an eclipsing binary inside the \corot\ PSF, the blend scenario having the highest probability is \targetc\ being a genuine transiting planet, while one of the five stars at about 10\arcsec\ is an adequate eclipsing binary. Using the same assumptions as above, we estimate the probability to have such a configuration as 3.3$\cdot$10$^{-5}$ (=5:130$\times$0.85$\cdot$10$^{-3}$). If the close targets are excluded as eclipsing binaries (either with a photometric campaign of with a few radial velocities taken in each of the stars), then the eclipsing binary should be non resolved in the NACO image, with a probability of $<$1:44000 = 2.3$\cdot$10$^{-5}$ estimated above. 
 
Taken together, the scenario in which one of the transiting signals is due to a planet is at least 966$\times$ times more plausible than a double-blend system. The scenario of a multiple transiting system in which both transit signals are due to planets is at least 26$\times$ more probable than a transiting planet + eclipsing binary blend (=0.85$\cdot$10$^{-3}$ / 3.3$\cdot$10$^{-5}$ ). This can rise to 37$\times$ if the five stars at a distance of about 10\arcsec\ from the target are definitely excluded as eclipsing binaries.

\subsection{Dynamical stability}

We calculated the orbital separation of \targetb\ and \targetc\ in Hill-units, to get an idea of the stability of the system. With the upper limit estimated for \targetb\ and the measurement of the mass of \targetc\, we estimate the orbital separations in Hill units of around eight. A two-planet system is supposed to be stable in Gyr time scales if this number is larger than seven \citep{funk2010}, so we conclude that dynamical stability is not a problem for the \target\ system.

\section{Conclusion}

The light curve of \target\ is compatible with the transits of two Neptune-sized planets (with radii of 3.7 and 5.0~\Rearth) at 5.11 and 11.76~d periods. We performed different photometric and spectroscopic measurements to try to secure the planetary nature of the objects and measure their masses and densities. The relative faintness of the target and the expected low amplitude of the radial velocity variations do not allow a precise mass measurement of \targetb\ and we provide an upper mass limit of 5.7~\Mearth, while for \targetc\ we obtain a tentative mass of 28$\pm$11~\Mearth. The deduced low densities, of $<$0.9\gcm3\ and 1.3$^{+0.5}_{-0.4}$\gcm3\ are indicative planets with significant gaseous envelopes.

We have studied the different configurations of four and five body systems that might reproduce the photometric light curve, which allowed a rejection of several configurations. A combination of these simulations with a study of the stellar number density in the direction of the target shows that the planetary scenario is at least 900 times more probable than a blend involving two eclipsing binary systems, and the multiple transiting planets scenario is at least 26 times more probable than a single planet + blend involving one eclipsing binary. In our reasoning, we have not considered other factors that might contribute to the multiple planet scenario, as the fact that there is a tentative detection of a $\sim$940~d, slightly eccentric $\sim$1.5~\Mjup\ object orbiting around the main target. As there is not enough statistics in the current sample of known multiplanetary systems, we have not used this information to favor or penalize the planet scenario.

Thus \target\ is most probably a multiple transiting system consisting of two Neptune-sized planets and a third more massive object at a larger orbital separation. They orbit a K1V star with a higher metallicity than the Sun, that shows photometric activity at a level of about 2\% with time scales of about 29~d. We interpret this as an indication of the stellar rotation period. The system is thus a good example of the Neptune-sized multiple transiting candidates found in the Kepler mission \cite{borucki2011}, and shows the intense ground-based effort needed to confirm these objects and measure their masses, due in part to the relative faintness, but also to the additional difficulty of having non-transiting or longer period planets in the system that complicate the analysis.

To proceed further, more precise light curves of the transiting objects would allow a narrower range of scenarios that might reproduce all the observables. A photometric search for deep eclipses in the companions at about 10\arcsec\ of the target should be performed with the ephemeris of the more challenging candidate \targetb. Alternatively, these companions could be detected spectroscopically. Despite being challenging for ground-based observations, the progress in the recent years have shown that it might be possible to detect the transits of \targetc\ (and probably also \targetb) and thus refine the ephemeris. 

A more intense radial velocity campaign devoted to \target\ might result in a more precise measurement of the masses of the planets, and thus their densities. These data could serve to better model the long period signal seen in the radial velocities, and to evaluate the effects of additional planetary companions in the system not considered in this work. Since the star shows photometric activity at the 2\% level and a period of around 29~days, the observing strategy should be designed in order to minimize or to understand better the effect of the activity on the measured masses. The star is currently the only multiple transiting system observable from several observatories at the southern hemisphere. Thus, \target\ could be a good target for future high resolution spectrographs planned for bigger aperture telescopes, such as ESPRESSO/VLT.

\begin{acknowledgements}

We wish to thank the ESO staff for their support and for their contribution to the success of the HARPS project and operation at ESO La Silla Observatory. Some of the data presented are from on observations made with the IAC80 operated at Teide Observatory on the
island of Tenerife by the Instituto de Astrof\'\i sica de Canarias.
The team at the IAC acknowledges support by grants
ESP2007-65480-C02-02 and AYA2010-20982-C02-02 of the Spanish Ministry
of Science and Innovation (MICINN). The CoRoT/Exoplanet catalogue (Exodat) was made possible by observations
collected for years at the Isaac Newton Telescope (INT), operated on the island
of La Palma by the Isaac Newton group in the Spanish Observatorio
del Roque de Los Muchachos of the Instituto de Astrof\'\i sica de Canarias.
The German CoRoT team (TLS and University of Cologne) acknowledges
DLR grants 50OW0204, 50OW0603, and 50QM1004. The French team wishes to
thank the CNES and the French National Research Agency (ANR-08- JCJC-
0102-01) for their continuous support to our planet search program. The Swiss team
acknowledges the ESA PRODEX program and the Swiss National Science
Foundation for their continuous support on CoRoT ground follow-up. R.A. acknowledges support from Spanish Ministry through its Ram\'on y Cajal program RYC-2010-06519.
A.S. Bonomo acknowledges a CNES grant. S. Aigrain acknowledges
STFC grant ST/G002266. M. Gillon acknowledges support from
the Belgian Science Policy Office in the form of a Return Grant. M. Endl,
W.D. Cochran and P.J. MacQueen were supported by NASA Origins of Solar
Systems grant NNX09AB30G. T. Mazeh acknowledges the supported of the Israeli Science Foundation (grant no. 655/07). A.S. acknowledges the support from the European Research Council/European Community under the FP7 through Starting Grant agreement number 239953. This research has made use of the SIMBAD database, operated at the CDS, Strasbourg, France, and of NASAÕs Astrophysics Data System. Part of the data presented herein were obtained at the W.M. Keck Observatory from telescope time allocated to the National Aeronautics and Space Administration through the agencyÕs scientific partnership with the California Institute of Technology and the University of California. The Observatory was made possible by the generous financial support of the W.M. Keck Foundation.

\end{acknowledgements}

\bibliographystyle{aa} 
\bibliography{references} 

\begin{table}
\caption{Radial-velocity measurements }
\begin{minipage}[t]{7.0cm} 
\begin{tabular}{llll}
\hline 
BJD-2,400,000.&  RV& $\sigma$&Instr\\
&[km/s]&[km/s]&\\
\hline
55170.975685 &  0.047  &  0.021  & HIRES\\
55221.936568 &  0.022  & 0.010   & HIRES\\
55222.971497 &  0.038  & 0.013   & HIRES\\
55224.880813 &  0.045  & 0.011   & HIRES\\
55610.830256 & -0.025  & 0.012   & HIRES\\
55610.999716 & -0.035  & 0.014   & HIRES\\
55934.798229 &  -0.001 &  0.006   & HIRES\\
55934.875484 &  -0.006  &  0.006   & HIRES\\
55935.069285 & -0.032  &  0.008   & HIRES\\
55935.811141 &   0.002  &  0.006   & HIRES\\
55935.912918 &    0.004 &   0.006  & HIRES\\
55936.047423 &  -0.018 &   0.007  & HIRES\\
55936.802950 &  -0.015 &   0.008  & HIRES\\
55936.921978 &  -0.016 &   0.006  & HIRES\\
55937.065640 &   0.028 &   0.009  & HIRES\\
55937.790470 &  -0.012 &   0.005  & HIRES\\
55937.973689 &  -0.020 &   0.004  & HIRES\\
55938.068929 &   -0.008 &   0.007  & HIRES\\
55961.765078 &   -0.004 &   0.006  & HIRES\\
55961.854540 &   -0.008 &   0.009  & HIRES\\
55961.978822 &   -0.005 &   0.005  & HIRES\\
55962.763358 &   -0.004 &   0.008  & HIRES\\
55962.934853 &   0.015 &   0.007  & HIRES\\
55963.894449 &    0.006 &   0.007  & HIRES\\
55964.761177 &  -0.010 &   0.007  & HIRES\\
55964.978007 &   0.014 &   0.005  & HIRES\\
55967.760449 &   -0.008 &   0.008  & HIRES\\
56314.788992 &    0.002 &   0.008  & HIRES\\
56314.883034 &  -0.024 &   0.010  & HIRES\\
56315.763203 &   -0.001 &   0.007  & HIRES\\
56315.863308 &   -0.001 &   0.005  & HIRES\\
56315.983394 &   0.012 &   0.009  & HIRES\\
56316.769601 &   -0.009 &  0.011  & HIRES\\
56316.990463 &    0.001 &  0.011  & HIRES\\
56317.764979 &    0.001 &   0.008  & HIRES\\
56317.875384 &   0.021 &  0.011  & HIRES\\
56317.989249 &    0.005 &  0.011  & HIRES\\
\hline
55163.78250	& 51.367 & 0.021 & HARPS\\
55168.80944	& 51.399 & 0.023 & HARPS\\
55238.60592	& 51.382 & 0.019 & HARPS\\
55238.65085	& 51.332 & 0.017 & HARPS\\
55239.58374	& 51.373 & 0.013 & HARPS\\
55239.62694	& 51.379 & 0.012 & HARPS\\
55244.63174	& 51.360 & 0.011 & HARPS\\
55244.67206	& 51.352 & 0.014 & HARPS\\
55873.82681	& 51.297 & 0.015 & HARPS\\
55887.75733 	& 51.315 & 0.028 & HARPS\\
55888.76728	& 51.361 & 0.024 & HARPS\\
55889.71362	& 51.302 & 0.048 & HARPS\\
55890.71149	& 51.330 & 0.021 & HARPS\\
55891.77165	& 51.321 & 0.016 & HARPS\\
55893.70971	& 51.319 & 0.020 & HARPS\\
55894.75240	& 51.301 & 0.013 & HARPS\\
55895.74860	& 51.346 & 0.028 & HARPS\\
55897.73358	& 51.367 & 0.017 & HARPS\\
55898.73570	& 51.365 & 0.021 & HARPS\\
55899.75383	& 51.321 & 0.014 & HARPS\\
55900.72193	& 51.315 & 0.030 & HARPS\\
55940.73120	& 51.314 & 0.021 & HARPS\\
55941.60339	& 51.363 & 0.017 & HARPS\\
55942.59996	& 51.311 & 0.030 & HARPS\\
55944.70908	& 51.321 & 0.021 & HARPS\\
55945.60067	& 51.379 & 0.022 & HARPS\\
55947.70024	& 51.344 & 0.015 & HARPS\\
55948.70078	& 51.312 & 0.013 & HARPS\\
55949.58968	& 51.346 & 0.023 & HARPS\\
55950.60120	& 51.327 & 0.018 & HARPS\\
55951.60073	& 51.337 & 0.017 & HARPS\\
56248.72977	& 51.349 & 0.031 & HARPS\\
56251.77125	& 51.344 & 0.015 & HARPS\\
56255.77752	& 51.328 & 0.016 & HARPS\\
\hline
\vspace{-0.5cm}
\end{tabular}
\end{minipage}
\label{rv}
\end{table}

\begin{table}[h]
\caption{ IDs, coordinates and magnitudes.}            
\centering        
\begin{minipage}[!]{7.0cm}  
\renewcommand{\footnoterule}{}     
\begin{tabular}{lcc}       
\hline\hline                 
CoRoT window ID & LRa02-E1-4601 \\
CoRoT ID & 300001097\\
USNO-B1 ID  & 0862-0103851\\
UCAC2 ID  & 30490990\\
UCAC3 ID & 173-47999\\
DENIS ID  & J064741.4-034309\\
2MASS ID   &  06474141-0343094\\
\\
\multicolumn{2}{l}{Coordinates} \\
\hline            
RA (J2000)  & 6h47m41.41s \\
Dec (J2000) &  -03$^{\circ}$43\arcmin09.5\arcsec \\
\\
\multicolumn{3}{l}{Magnitudes} \\
\hline
\centering
Filter & Mag & Error \\
\hline
r'$^a$ & 15.113 & --\\
i'$^a$ & 14.305 & 0.05 \\
J$^b$  & 13.595 & 0.021 \\
H$^b$  & 13.046 & 0.03\\
K$^b$  & 12.924& 0.023\\
\hline\hline
\vspace{-0.5cm}
\footnotetext[1]{Provided by Exo-Dat (Deleuil et al, 2008);}
\footnotetext[2]{from 2MASS catalog.}
\end{tabular}
\end{minipage}
\label{startable}      
\end{table}

\begin{table*}
\centering
\caption{Planet and star parameters.}            
\begin{minipage}[t]{13.0cm} 
\setlength{\tabcolsep}{10.0mm}
\renewcommand{\footnoterule}{}                          
\begin{tabular}{l l}        
\hline\hline                 
\\
\multicolumn{2}{l}{\emph{Ephemeris}} \\
\hline
Planet $b$ orbital period $P$ [days] &  5.1134 $\pm$ 0.0006  \\
Planet $b$ transit epoch of  $T_{tr_b}$ [HJD-2400000] & 54789.611 $\pm$ 0.006 \\
Planet $b$ transit duration $d_{tr}$ [h] & 2.85 $\pm$ 0.25 \\
\hline
Planet $c$ orbital period $P$ [days] & 11.759 $\pm$ 0.0063 \\
Planet $c$ transit epoch $T_{tr_c}$ [HJD-2400000] & 54795.3803 $\pm$ 0.0265  \\
Planet $c$ transit duration $d_{tr}$ [h] & 4.8 $\pm$ 0.2 \\
\\
\multicolumn{2}{l}{\emph{Results from radial velocity observations}} \\
\hline    
Radial velocity semi-amplitude $K_b$ [\ms] & $<$2.1 \\
Radial velocity semi-amplitude $K_c$ [\ms] & 8.3$^{+2.9}_{-2.9}$ \\
Systemic velocity  $V_{r}$ [\kms] & $\sim$51.37\\
Eccentricity $b$ and $c$ & 0. (fixed)\\
\\
\multicolumn{2}{l}{\emph{Fitted transit parameters}} \\
\hline
Radius ratio $k_b=R_{p}/R_{*}$ & 0.0370 $\pm$ 0.0021\\
Radius ratio $k_c=R_{p}/R_{*}$ & 0.0500 $\pm$ 0.0010\\
Inclination $i_b$ [deg] & 86.5 $\pm$ 2\\
Inclination $i_c$ [deg] & 89 $^{+1}_{-1.6}$ \\
Phase of transit ingress $\theta_{1,b}$ & 0.0116 $\pm$ 0.0010 \\
Phase of transit ingress $\theta_{1,c}$ & 0.0084 $\pm$ 0.0003 \\
\\
\multicolumn{2}{l}{\emph{Adopted parameters}} \\
\hline
Limb darkening coefficient $u_a$ & 0.58 $\pm$ 0.04 \\
Limb darkening coefficient $u_b$ & 0.13 $\pm$ 0.04 \\
\\
\multicolumn{2}{l}{\emph{Deduced transit parameters}} \\
\hline
Scaled semi-major axis $a_{b}/R_{*}$ & 10.9 $\pm$ 2.8 \\
Scaled semi-major axis $a_{c}/R_{*}$  & 18.0 $\pm$ 2.5 \\
$M^{1/3}_{*}/R_{*}$ [solar units], estimated from transits of $b$ & 0.88 $\pm$ 0.22 \\
$M^{1/3}_{*}/R_{*}$ [solar units], estimated from transits of $c$ & 0.85 $\pm$ 0.13 \\
Impact parameter $b_b$ & 0.65 $\pm$ 0.3\\
Impact parameter $b_c$ & 0.3 $\pm$ 0.4\\
\\
\multicolumn{2}{l}{\emph{Spectroscopic parameters }} \\
\hline
Effective temperature $T_{eff}$[K] & 4950 $\pm$ 150\\
Surface gravity log\,$g$ [dex]&  4.55 $\pm$ 0.15 \\
Metallicity $[\rm{Fe/H}]$ [dex]&  0.30 $\pm$ 0.15\\
Stellar rotational velocity {\vsini} [\kms] & 2.0$^{+1.0}_{-1.5}$ \\
Spectral type & K1V\\
\\
\multicolumn{2}{l}{\emph{Stellar and planetary physical parameters from combined analysis}} \\
\hline
Star mass [\Msun] & 0.91 $\pm$ 0.09\\
Star radius [\Rsun] &  0.86 $\pm$  0.09\\
Distance of the system [pc] & 600 $\pm$ 70\\
Interstellar extinction [mag] & 0.30 $\pm$ 0.25 \\
Stellar rotation period $P_{rot}$ [days]  &   $\sim$29 \\
Age of the star $t$ [Gyr] & $\sim$11 \\
Orbital semi-major axis $a_b$ [AU] & 0.056$\pm$ 0.002\\
Orbital semi-major axis $a_c$ [AU] & 0.098$\pm$ 0.003\\
Planet $b$ mass $M_{p}$ [M$_J$ ]$^a$ &   $<$ 0.018 \\
Planet $c$ mass $M_{p}$ [M$_J$ ]$^a$ &   0.088$^{+0.035}_{-0.035}$ \\
Planet $b$ mass $M_{p}$ [M$_E$ ]$^a$ &   $<$ 5.7\\
Planet $c$ mass $M_{p}$ [M$_E$ ]$^a$ &   28$^{+11}_{-11}$ \\
Planet $b$ radius $R_{p}$[R$_J$]$^a$  &  0.33 $\pm$ 0.04 \\
Planet $c$ radius $R_{p}$[R$_J$]$^a$  &  0.44 $\pm$ 0.04 \\
Planet $b$ radius $R_{p}$[R$_E$]$^a$  &  3.7 $\pm$ 0.4 \\
Planet $c$ radius $R_{p}$[R$_E$]$^a$  &  5.0 $\pm$ 0.5 \\
Planet $b$ density $\rho_{p}$ [$g\;cm^{-3}$] & $<$ 0.9 \\
Planet $c$ density $\rho_{p}$ [$g\;cm^{-3}$] &  1.3 $^{+0.5}_{-0.4}$\\
Equilibrium temperature $^b$ planet $b$, $T_{eq,b}$ [K] & 1070 $\pm$ 140 \\
Equilibrium temperature $^b$ planet $c$, $T_{eq,c}$ [K] & 850 $\pm$ 80 \\
\\
\hline       
\vspace{-0.5cm}
\footnotetext[1]{Radius and mass of Jupiter taken as 71492 km and 1.8986$\times$10$^{30}$ g, respectively.}
\footnotetext[2]{Zero albedo equilibrium temperature for an isotropic planetary emission.} 
\end{tabular}
\end{minipage}
\label{starplanet_param_table}  
    
\end{table*}

\end{document}